# Estimating the Roles of Protonation and Electronic Polarization in Absolute Binding Affinity Simulations

*Edward King*[1], *Ruxi Qi*[4], *Han Li*[2], *Ray Luo*[1,2,3*], *and Erick Aitchison*[1*]

[1]Departments of Molecular Biology and Biochemistry, [2]Chemical and Biomolecular Engineering, [3]Materials Science and Engineering and Biomedical Engineering, University of California, Irvine, California 92697 (USA)
[4]Cryo-EM Center, Southern University of Science and Technology, Shenzhen, Guangdong 518055 (China)

**Abstract**

Accurate prediction of binding free energies is critical to streamlining the drug development and protein design process. With the advent of GPU acceleration, absolute alchemical methods, which simulate the removal of ligand electrostatics and van der Waals interactions with the protein, have become routinely accessible and provide a physically rigorous approach that enables full consideration of flexibility and solvent interaction. However, standard explicit solvent simulations are unable to model protonation or electronic polarization changes upon ligand transfer from water to the protein interior, leading to inaccurate prediction of binding affinities for charged molecules. Here, we perform extensive simulation totaling ~540 µs to benchmark the impact of modeling conditions on predictive accuracy for absolute alchemical simulations. Binding to urokinase plasminogen activator (UPA), a protein frequently overexpressed in metastatic tumors, is evaluated for a set of ten inhibitors with extended flexibility, highly charged character, and titratable properties. We demonstrate that the alchemical simulations can be adapted to utilize the MBAR/PBSA method to improve the accuracy upon incorporating electronic polarization, highlighting the importance of polarization in alchemical simulations of binding affinities. Comparison of binding energy prediction at various protonation states indicates that proper electrostatic setup is also crucial in binding affinity prediction of charged systems, prompting us to propose an alternative binding mode with protonated ligand phenol and Hid-46 at the binding site, a testable hypothesis for future experimental validation.

**Introduction**

Electrostatics and polarization effects are critical to the study of biomolecular processes such as dynamics, recognition, and enzymatic catalysis. The success of computational simulation in sampling physiologically apt biomolecular structures involved in enzyme activity is dependent on both efficient calculations to enable consideration of atomic interactions at long timescales, and accurate treatment of those interactions to maximize predictive capability. Current simulation efforts often ignore the impact of electronic polarization due to their complexity and high computational costs, leading to errors such as the overestimation of gas-phase water dimer interaction energy by greater than 30% with the nonpolarizable TIP5P model.[1,2] The standard nonpolarizable point-charge model allows analysis of electrostatics through straightforward application of the Coulombic potential but is unable to capture the effect of exposure to different electrostatic environments such as between the protein interior and solvent that is essential to biomolecular processes. Furthermore, reference parameters for nonpolarizable models are typically derived from gas-phase quantum mechanical calculations, resulting in spurious "pre-polarization" when used in an aqueous environment due to the inclusion of average bulk polarization effects inconsistent with the liquid phase. Improving the treatment of electrostatics and polarization would significantly enhance efforts to study the biomolecular processes of ion-dependent interactions, proton and electron transfer in enzyme catalysis, order-disorder transitions in intrinsically disordered regions, pKa effects in titration, etc.

A number of polarizable models have been developed to address the accurate representation of electrostatic interactions for biomolecular simulation including the OPLS-AA fluctuating charge model[3, 4], Drude oscillator[5-8] with CHARMM, and AMOEBA with multipole expansion and increased force field components.[9-11] Recent developments with AMBER include the polarizable Gaussian Multipole (pGM)[12, 13] model that improves over the previous induced dipole implementation based on Thole models.[14-18] pGM represents each atom's multipole as a single Gaussian function or its derivatives, speeding electrostatic calculations over alternative Gaussian-based models. By screening short-range interactions in a physically consistent manner, pGM enables the stable charge-fitting necessary to describe molecular anisotropy that is difficult to achieve with Thole models.[12, 13]

Regardless of which model to use, an important application of molecular simulations is the accurate prediction of binding affinities to accelerate the drug discovery process as recently reviewed.[19] Accurate virtual screening is necessary to reduce the excessive time and costs associated with drug development, which are estimated to be over 10 years and $2.8 billion for an approved drug.[20] Methods based on geometric docking to optimize the shape and electrostatic complementarity between binding partners,[21-25] end-point MD simulations with either the linear interaction energy method[26] or the Molecular Mechanics Poisson Boltzmann Surface Area method,[27-36] end-point MC simulations with the Mining Minima method,[37-40] alchemical pathway simulations with full sampling of conformational flexibility in explicit solvent,[22, 41-47] and machine learning based on correlation of structural features and protein-ligand interactions[48-50] have shown promise, but have not achieved the generalizable accuracy required or come at too high computational cost for practical application to drug discovery.

Alchemical simulations measure the free energy difference between two states, so that it can be used to determine the free energy change between the complex state with protein and ligand bound and the unbound state with protein and ligand separated.[51] Alchemical simulations progress through a closed thermodynamic cycle, utilizing transformations through unphysical intermediate states modeling the gradual decoupling of ligand electrostatic and van der Waals (VDW) interactions with the protein environment, and provide a computational advantage over brute-force simulations of unbinding or binding processes.[52] Previous work has highlighted the utility of alchemical simulations in the computation of small molecule distribution coefficients between solvent phases,[53] protein stability upon amino acid mutation,[54] binding affinity through relative transformation growing or deleting functional groups off a reference structure,[55-57] and absolute transformation where the larger perturbation of ligand transfer to gas phase is modeled.[58-62] Absolute alchemical transformations, which permit direct prediction of binding energy and do not require initialization from a reference structure with high similarity to the target as relative calculations, have only recently become practical with the development of high-performance computer hardware, such as graphical processing units (GPUs).

Structure-based drug design coupled to alchemical simulations has served as the foundation for drug development campaigns;[58] however, limitations due to heterogeneity in protocols and model setups, limited accuracies in molecular force fields, and insufficient sampling of the protein and ligand conformations still impede prediction accuracy. Furthermore, standard alchemical simulations are unable to model protonation or electronic polarization changes upon ligand transfer from water to the protein interior, leading to inaccurate prediction of binding affinities for charged molecules. In this study, we benchmarked the absolute alchemical transformation methods on the urokinase plasminogen activator (UPA) system to estimate the impacts of protonation and closely related polarization effects during the protein-ligand binding process. UPA is a serine protease that activates plasmin which is involved in the degradation of blood clots and extracellular matrix.[63] UPA has been found to be overexpressed in several types of metastatic tumors; this upregulation has been proposed to drive the tissue degradation required for cancer invasion and metastatic growth, making UPA a desirable target for anticancer therapeutics. The tested models are a set of high-resolution crystal structures collected by Katz et al. with 10 different competitive inhibitors of varying sizes, charges, and chemical groups (Figure 1).[64-66] The inhibition constant ($K_i$) of each ligand has been experimentally determined, allowing for the validation of our computational protocols. This set of ligands represents a diverse and challenging test case with inhibitors bearing a large number of torsion angles that require lengthy simulation to sample the available conformation space, highly charged character amplifying inaccuracy in the treatment of electronic polarization, and multiple potential protonation states due to ionizability and tautomerization.

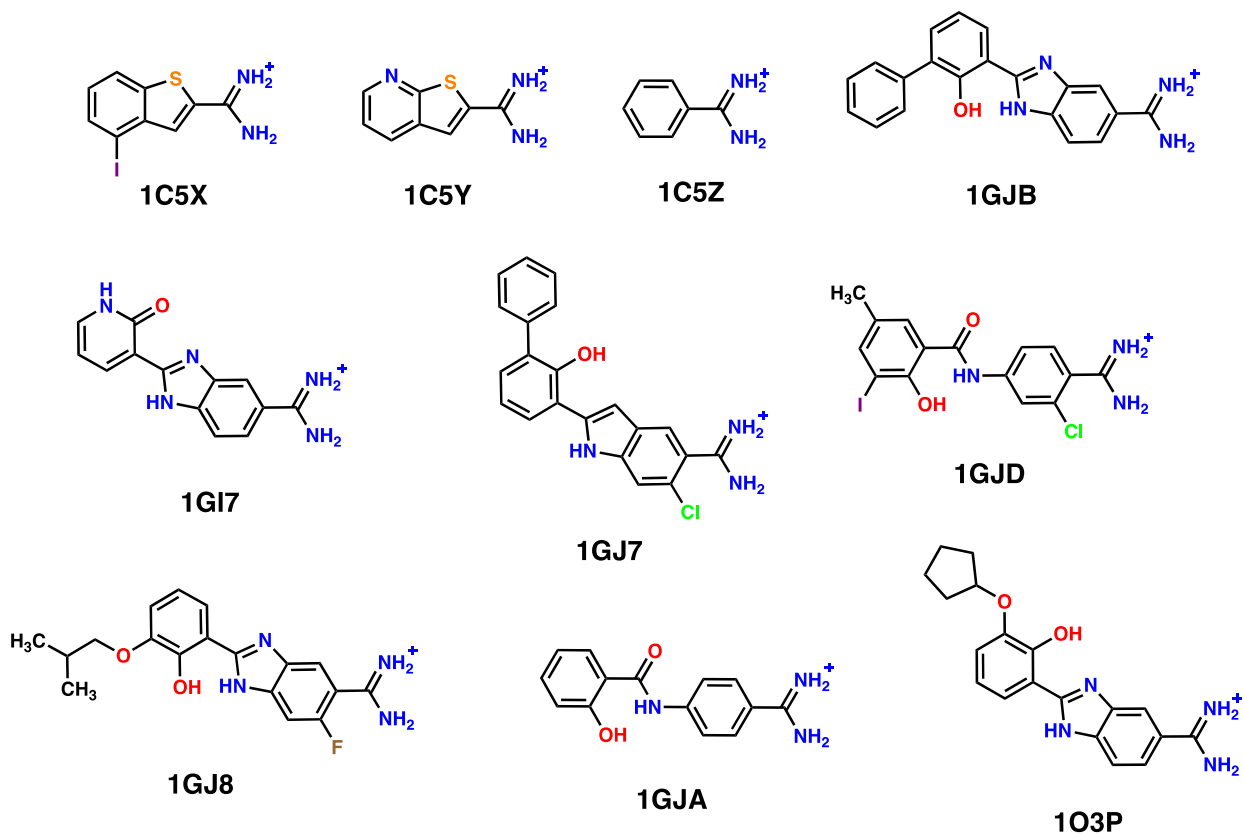

**Figure 1. Chemical structures of the 10 evaluated UPA inhibitors.** The molecules share a benzamidine-like scaffold with characteristic amidine group carrying positive charge, and extended tails comprised of a phenol group and other functional modifiers. The hydroxyl on the phenol is proposed to be titratable and samples deprotonated and protonated states during binding, altering the hydrogen bonding capability of the ligands. The inhibitors are categorized as small (those without the phenol group): 1C5X, 1C5Y, 1C5Z, and 1GI7, and big for those with potentially charged phenols: 1GJ7, 1GJ8, 1GJA, 1GJB, 1GJD, 1O3P.

To address the heterogeneity in the alchemical protocol, we studied the effects of various simulation setups/conditions including ligand force field choices, salt concentrations, alternative protonation states, and ligand restraints through a single harmonic distance (1DOF) or with the more involved 6-degree of freedom (6DOF) restraints to improve convergence by maintaining the ligand in the binding pose. We further developed a new strategy to utilize the PBSA continuum solvent model coupled with the Multistate Bennet Acceptance Ratio (MBAR) approach to estimate the effect of electronic polarization in this challenging set of highly charged ligands. We demonstrated that the application of the MBAR/PBSA method with optimized solute dielectric constant permits more properly modeled electronic polarization, leading to superior accuracy in the absolute binding free energy prediction of this highly charged set of ligands. This allows us to assess alternative protonation states for the ligands and titratable residues in the binding pose, offering a testable hypothesis for future experimental validation.

## Methods

Structure Preparation for Molecular Simulations

Crystal structures for the inhibitor bound urokinase plasminogen activator (PDB: 1C5X, 1C5Y, 1C5Z, 1GI7, 1GJ7, 1GJ8, 1GJA, 1GJB, 1GJD, and 1O3P)[64-66] were obtained from the RCSB PDB database.[67] Experimentally determined binding free energies were obtained from the PDBbind database.[68] The structures were prepared for simulation by removal of all water molecules greater than 5 Å away from the active site, removal of all co-crystallized ligands that were not the target inhibitor, and truncation of all structures to 245 amino acids by deletion of disordered C-terminal residues that were not resolved in all crystal structures (a maximum of 3 residues were deleted). Disulfide bonds were added as in the crystal structures.

As there is no pKa measurement available, protonation states of titratable residues were determined at pH 7.4 through the H++ webserver[69] except those in the binding pocket, where the complex crystal structures were used to infer the likely protonation states. In the binding pocket, His-94 and Asp-97 are modeled as default neutral HIE and charged ASP, respectively, as there is no unusual polar interaction with the ligand molecule. For His-46 and the ligand phenol group, there are two possible protonation states that satisfy the steric constraint in the crystal structures as discussed in detail in Results and Discussion. The first possibility is to set His-46 as protonated HIP and the ligand phenol as deprotonated as suggested in Ref.[65] The second possibility is to set His-46 as deprotonated HID and the ligand phenol as protonated. Given these protonation states, 1C5X, 1C5Y, 1C5Z, and 1GI7 are treated as +1 net charge due to protonation at the amidine group, and all other ligands are treated as +0 net charge zwitterions with a +1 charge of the amidine group and -1 charge on the deprotonated phenol hydroxyl or as +1 net charge ions with 0 charge on the protonated phenol hydroxyl. The predictive accuracy of both protonation states was compared to a baseline model with all ligands treated as +1 net charge due to default protonation at the amidine and phenol hydroxyl with Hip-46. All tested conditions and protonation states are summarized in Table 1.

| Condition | HIS46 | Protonated Ligands (+1 charge) | Deprotonated Ligands (+0 charge) | Salt | Restraint Potential |
|---|---|---|---|---|---|
| Baseline | HIP | 1C5X, 1C5Y, 1C5Z, 1GI7, 1GJ7, 1GJ8, 1GJA, 1GJB, 1GJD, 1O3P | - | Counter-ions only | 1DOF |
| Baseline + 150 mM salt | HIP | 1C5X, 1C5Y, 1C5Z, 1GI7, 1GJ7, 1GJ8, 1GJA, 1GJB, 1GJD, 1O3P | - | 150mM | 1DOF |
| Baseline + Deprotonated Ligands | HIP | 1C5X, 1C5Y, 1C5Z, 1GI7 | 1GJ7, 1GJ8, 1GJA, 1GJB, 1GJD, 1O3P | Counter-ions only | 1DOF |
| All-HIP | HIP | 1C5X, 1C5Y, 1C5Z, 1GI7 | 1GJ7, 1GJ8, 1GJA, 1GJB, 1GJD, 1O3P | 150 mM | 1DOF/6DOF |
| All-HID | HID | 1C5X, 1C5Y, 1C5Z, 1GI7, 1GJ7, 1GJ8, 1GJA, 1GJB, 1GJD, 1O3P | - | 150 mM | 1DOF |
| Small-HIP | Mixed | 1C5X, 1C5Y, 1C5Z, 1GI7 (HIP) 1GJ7, 1GJ8, 1GJA, 1GJB, 1GJD, 1O3P (HID) | - | 150 mM | 1DOF |
| Small-HID | Mixed | 1C5X, 1C5Y, 1C5Z, 1GI7 (HID) | 1GJ7, 1GJ8, 1GJA, 1GJB, 1GJD, 1O3P (HIP) | 150 mM | 1DOF |

**Table 1. Summary of simulation conditions.** The baseline corresponds to a default setup with full ligand and protein protonation, salt concentration at charge neutralizing amount, and 1DOF restraint. Singular condition changes to the baseline: 150 mM salt concentration, and deprotonated ligand phenol. Alternative protonation states are tested with variable ionization at the ligand phenol and His-46 to model the effect of hydrogen bonding potential on binding free energy prediction.

Ligand partial charges were determined with the Restrained Electrostatic Potential (RESP) method[70] at the HF/6-31G* level using Gaussian09,[71] except HF/CEP-31G was used for ligands with iodine. Other ligand parameters were taken from the General Amber Force Field (GAFF)[72] or GAFF2. The protein was modeled with the ff14sb[73] force field. Systems were solvated in TIP3P[74] water in truncated octahedron with 10 Å buffer, and charge neutralized with $Na^+$/$Cl^-$ ions. Additional ions were also added to reach 150 mM salt concentration under the high salt condition tested. Molecular dynamics simulations were performed with pmemd.cuda[75] from the Amber18 package with an 8 Å Particle Mesh Ewald[76] cutoff and otherwise default settings.

Alchemical Simulation Protocol

Computation of binding free energies was conducted through a four-step process: equilibration, restraint sampling, decharging, and softcore van der Waals removal.[77, 78] Imposition of restraints and each of the two inhibitor transformation steps (decharging and VDW removal) proceeded through a series of alchemical intermediates described by the coupling parameter lambda increasing from 0 (starting state) to 1 (fully transformed ending state). Final simulation data are an aggregate of ensemble MD of five independent replicates started from the minimized crystal structures with randomized initial velocities. The free energy differences between states were calculated with MBAR[79, 80] through the pymbar[79] package, and required the calculation of energy cross-terms for each trajectory at each restraint, charge, and VDW lambda steps. Only data produced from frames in the last half of each trajectory were included in energy calculations to ensure well-equilibrated results.

*Minimization and Equilibration* The UPA systems were minimized in two steps: first with 2,500 steps of steepest descent and 2,500 steps of conjugate gradient where all non-hydrogen solute atoms were restrained with a 20 kcal mol$^{-1}$ Å$^{-2}$ force to relieve steric clash. The second minimization to remove solute steric clashes was run with the same cycle settings and restraints removed. Heating from 0 K to 298 K was performed over 0.5 ns with 10 kcal mol$^{-1}$ Å$^{-2}$ restraints on all non-hydrogen solute atoms. Solvent density equilibration under the NPT condition and the Langevin thermostat with collision frequency 2 ps$^{-1}$ was carried out over 0.4 ns with 2 kcal mol$^{-1}$ Å$^{-2}$ restraints on all non-hydrogen solute atoms to stably reach 1 atm pressure. Next, an unrestrained 100 ns NVT equilibration with the Langevin thermostat and collision frequency 1 ps$^{-1}$ was completed to clear remaining structural artifacts from the initial crystal structure. Separate simulations for the unrestrained inhibitor alone and the protein-inhibitor complex were run for the discharging and VDW removal process. The inhibitor alone was extracted from the equilibrated complex and solvated in the TIP3P truncated octahedron box with 20 Å buffer and neutralized with Na$^+$/Cl$^-$ counter-ions or up to 150 mM salt concentration. Trajectory data was analyzed with the cpptraj program[81] and the NumPy[82] packages.

*Imposing Restraints* As electrostatic and VDW interactions are decoupled, the ligand has the ability to escape the active site and sample states irrelevant to binding, hindering convergence. Standard practice is to apply a restraint on the ligand which requires calculating the free energy contribution of the restraints,

$$\Delta A_r = -kT \ln \frac{Z_P Z_L}{Z_{CL}}, \tag{1}$$

where $Z_P$, $Z_L$, $Z_{CL}$ are the configurational partition functions of the protein, ligand, and the cross-linked state.[83-87] The derivation of the restraint free energy depends on the external degrees of freedom restrained on the ligand relative to the protein, which defines the cross-linked state or virtual bond. Since the ligand position and/or orientation is restrained to the protein, the protein external degrees of freedom can be separated out leaving the integration of the internal and external degrees of freedom for the ligand. The restraint free energy can be simplified into the difference between the term from the integration of all external degrees of freedom of a non-linear ligand, $8\pi^2 V$, and the term of a gaussian integral for each degree of freedom used to restrain the ligand, $\sqrt{\frac{2\pi k_b T}{K_\xi}}$, where $K_\xi$ is the harmonic restraint force constant.

In the 1DOF restraint, a single harmonic distance restraint with a 20 kcal mol$^{-1}$ Å$^{-2}$ force constant was utilized as a virtual bond between the Asp-192 alpha carbon and the ligand amidine carbon. The final analytical correction for the single distance restraint is as follows for restraining the ligand in the unbound state,[84, 86]

$$-k_b T \ln \left[ \frac{8\pi^2 V^0 K_r^{1/2}}{(2\pi k_b T)^{1/2}} \right] \tag{2}$$

where $K_r$ is the force constant of the distance restraint (SI Figure 1) and $V^0$ is the standard state volume. This virtual bond restraint is relative to the protein. This is different from the cartesian position restraint from Roux et al.,[83] which uses a point in three-dimensional cartesian space to restrain the ligand, resulting in an integral of $\left(\frac{2\pi k_b T}{K_\xi}\right)^{3/2}$.

To study the effects of the restraining protocol, an independent set of simulations was also run with the set of 6 degrees of freedom (6DOF) orientational restraints proposed by Boresch et al.[86] based on a single distance, two angular, and three dihedral parameters, all with 10 kcal mol$^{-1}$ Å$^{-2}$ force constants. For the 6DOF restraint, the final analytical correction for restraining the ligand in the unbound state is[86]

$$-k_b T \ln \left[\frac{8\pi^2 V^0 \left(K_r K_{\theta_A} K_{\theta_B} K_{\phi_A} K_{\phi_B} K_{\phi_C}\right)^{1/2}}{r_{a,A,0}^2 \sin \theta_{A,0} \sin \theta_{B,0} (2\pi k_b T)^3}\right] \quad (3)$$

where $r_{a,A,0}$ is the restrained distance, $\theta_{A,0}$ and $\theta_{B,0}$ are the two restrained angles and $K$'s are the force constants (SI Figure 1).

All restraint bounds were selected based on the final positions of the ligands at the end of the equilibration stage. Restraint sampling from off to full strength was performed over 6 equally spaced lambda values (0, 0.2, 0.4, 0.6, 0.8, 1.0), each with 10 ns. A separate analytical correction is calculated to determine the penalty for restraining the ligand in the unbound state.

*Alchemical Simulations* Decharging through parameter-interpolation of the inhibitors' partial charges to the sampled lambda window was performed to gradually decouple all electrostatic interactions between the inhibitor and environment and was separately run prior to VDW removal to avert the possibility of attractive atom overlap singularities. Decharging for both ligands alone and complex was performed linearly over 11 equally spaced lambda values (0.0, 0.1, 0.2, 0.3, 0.4, 0.5, 0.6, 0.7, 0.8, 0.9, 1.0), each for 40 ns with full restraints. System neutrality was maintained with charged ligands by simultaneously decharging a counter-ion alongside the ligand. Energies from lambda dependent VDW removal were calculated with the softcore potential to avoid numerical instability at endpoint lambdas observed with linear scaling due to atomic overlap.[77, 78] VDW removal was completed over 16 lambda values (0.0, 0.1, 0.2, 0.3, 0.4, 0.5, 0.55, 0.60, 0.65, 0.70, 0.75, 0.80, 0.85, 0.90, 0.95, 1.0) each for 20 ns, with denser sampling of lambda values at the later stages to more smoothly decouple VDW interactions. The dummy counter-ions present with charged inhibitor systems are VDW decoupled concurrently with the ligand. All free energy simulations were conducted with the pmemd.GTI[44] program in Amber18.

Alchemical simulation results for each ligand were aggregated from 5 individual trajectories with randomized starting velocities to ensure robust conformational sampling. Energy values from the last half of each lambda window for the replicates are concatenated together to combine equilibrated data for final MBAR analyses. Examination of convergence involves calculating the difference in final free energy with the addition of each replicate trajectory, the analysis shows that cumulative free energies with 5 replicates leads to less than 0.5 kcal/mol deviations. Achieving reasonable convergence in absolute binding affinities for the systems studied here was not trivial, and the total MD simulation time including equilibration, restraint sampling, decharging, and VDW removal was 1.7 µs for a single sample. The total cumulative MD simulation time including all tested conditions and replicates was ~540 µs. Raw traces of the changes in free energy with each lambda window illustrate the linearity of the decharging process, and the high variation of the VDW removal process in the simulation of the complex (SI Figures 2, 3, 4, 5).

Estimation of Electronic Polarization with MBAR/PBSA

The Poisson Boltzmann Surface Area (PBSA) method differs from the standard MD approach in that solvent molecules are modeled implicitly as a continuum in a mean-field manner rather than as explicit molecules, which offers significant simulation efficiency.[88-110] PBSA coupled with the MBAR protocol (i.e. MBAR/PBSA) was developed as an alternative to computing the decharging free energy for alchemical simulations in explicit solvent.[111] The original explicit solvent trajectories for all lambda windows used for decharging were prepared by stripping the waters and ions, but with the counter-ion used to maintain the ligand charge neutrality kept.

MBAR/PBSA energy evaluation was performed via the linear Poisson-Boltzmann (LPB) method with the Amber18 sander module[112] by post-processing solvent-stripped snapshots from alchemical simulations. Nonpolar solvation free energies[113] were turned off as only electrostatic interactions with and without polarization were compared. Following calculation of electrostatic free energies from the individual snapshots, the MBAR method was used to determine the composite free energy change for the complete decharging process as in the explicit solvent model. The PBSA parameters were set to 0.5 Å grid spacing with different interior dielectric constants ranging from the default of 1 to 2 and solvent dielectric constant 80. Periodic boundary conditions were used, and the box size was set to twice the size of the complex dimension or four times the size of the ligand dimension. The incomplete Cholesky conjugate gradient numerical LPB solver was utilized and the iteration convergence criterion set as $10^{-3}$.[114-116] Atomic radii were based on the default mbondi parameters in the Amber package.[112] The solvent probe radius was set to the default 1.4 Å and the mobile ion probe radius for the ion accessible surface was also set to the default 2.0 Å. The short-range pairwise charge-based interactions were cutoff at 7 Å, and long-range interactions were calculated from the LPB numerical solution.[117] Ionic strength was set to match the value from the explicit solvent MD simulations.

The solvation free energies computed from the PBSA model are critically dependent on the atomic radii. The Amber default mbondi radii parameters are revised from the Bondi radius set, and do not reproduce the solvation free energies with the TIP3P water as used in this study. Thus, the binding free energies from the explicit solvent trajectories were first utilized to calibrate the PBSA model through scaling of the ligand and protein radii at solute dielectric constant 1 to match the explicit solvent simulations as previously developed for free energy simulations of ionic systems.[111] First, ligand radii were uniformly scaled by the "Radiscale" PBSA input value and were tuned to minimize the absolute deviation between PBSA and explicit-solvent electrostatic free energies for the ligand alchemical simulations. Next given optimized "Radiscale", the protein radii were then uniformly scaled by the "Protscale" input value and were tuned to minimize the absolute deviation between PBSA and explicit-solvent electrostatic free energies for the complex alchemical simulations. Following calibration of the atomic radii, the PBSA model can be appropriately utilized for the investigation of electronic polarization by varying the solute dielectric constant.

**Results and Discussion**

Structural Agreement between Simulation and Experiment

Errors or deficiencies in sampling experimentally relevant conformations are attributed to standard MD protocols/force fields and highlight sources for inaccuracy in the downstream alchemical process that is sensitive to sampled conformations. Thus, we first analyzed the effects of force field choices on the quality of sampled conformations of UPA inhibitors prior to alchemical simulations. Here General Amber Force field (GAFF) and GAFF2 were both studied.

The 10 co-crystalized inhibitors share a common amidine group attached to an aromatic ring (Figure 1). The larger ligands maintain the benzamidine scaffold of the small ligands linked to a phenol-like ring and additional functional groups including methyl and cyclic structures, including 1GJ7, 1GJ8, 1GJA, 1GJB, 1GJD, 1O3P (termed big ligands below). The rest of the ligands, 1C5X, 1C5Y, 1C5Z, and 1GI7 are categorized as small ligands (1GI7 is large in size but lacks the characteristic phenol-like ring of the larger ligands so is grouped here).

Binding is mediated by two sets of polar interactions. One is from the positively charged amidine group, which is common among these inhibitors and makes a dense network of stabilizing polar contacts to a buried Asp-192 and Ser-193 in the active site (Figure 2A). The phenol hydroxyl makes an additional group of hydrogen bonds centered on Ser-198; worth noting is its short hydrogen bond to Ser-198 with distance ~2.2 Å, the lower bound of a hydrogen bond length (Figure 2B).[65] Due to the short distance, the phenol hydroxyl is inferred to act as an acid and be deprotonated in the bound state, and His-46 is interpreted to be a fully protonated Hip-46 to function as a hydrogen bond donor for the ligand phenol.[65] Interestingly, the hydrogen bond between the phenol hydroxyl and His-46 is longer at ~2.7 Å even if donor and acceptor are both charged when inferred this way. An alternative solution that satisfies the similar steric constraint in the crystal structure is for the protonated phenol hydroxyl to form the hydrogen bond with the Hid-46. In doing so, both groups are neutral. It should be pointed out that there is no direct pKa measurement of these residues/functional groups. Several of the inhibitors contain halogens (1C5X, 1GJ7, 1GJD, 1GJ8), which are not parameterized comprehensively in current force fields, possibly leading to inaccurate treatment of these ligands.

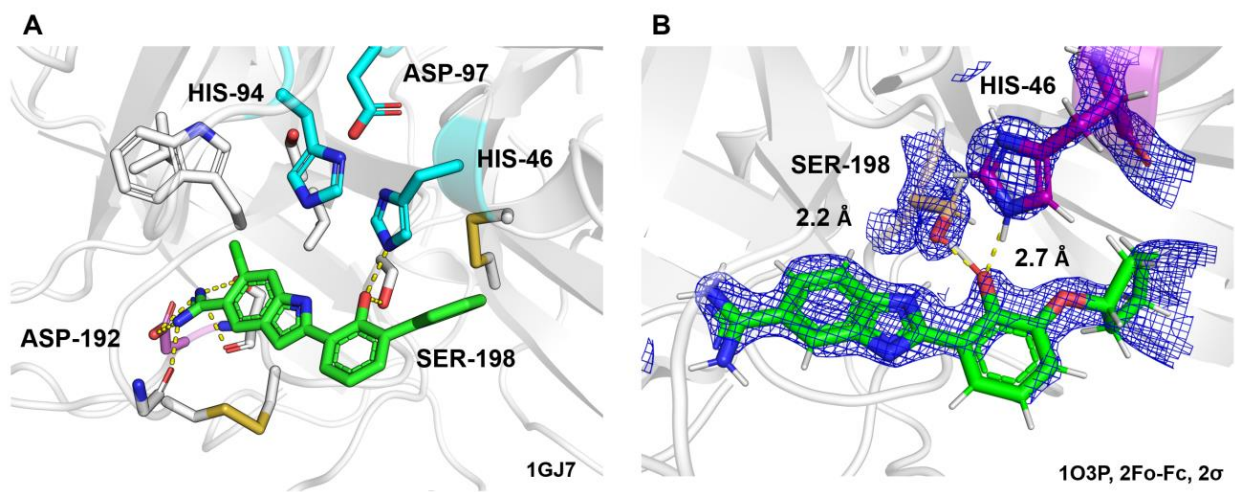

**Figure 2. Example inhibitor binding poses.** A) The protein and ligand form a network of polar interactions at two locations, at the base of the active site between the negatively charged Asp-192 and the positively charged amidine, and near the phenol hydroxyl with Ser-198 and His-46. B) Electron density supports the positioning of the ligand hydroxyl unusually close to Ser-198. An exceptionally short hydrogen bond is formed between the phenol hydroxyl and Ser-198 hydroxyl with distance ~2.2 Å, this interaction may not be captured accordingly with typical force fields due to van der Waals repulsion.

*Crystal Structure Analysis* The stability of binding sites and ligands in these crystal structures are evaluated through B-factor and electron density analyses. The binding pocket is defined to include any residue with atoms within 6 Å of the inhibitor (SI Figure 6). Crystal B-factors describing flexibility are not directly comparable between structures since they are a function of the crystalline disorder and resolution. Thus, the B-factors are normalized within each structure and Z-scores are compared (SI Figure 7). The binding pockets exhibit roughly equivalent stability with median B-factor Z-scores around -0.6 and the ligands mostly fall into the range of -0.5 to 0. The most stable ligand is 1GJB, possibly due to hydrophobic packing of the highly nonpolar and compact benzene tail, and the most flexible is 1GI7, which is large but lacks the phenol hydroxyl that enables the hydrogen bonding array at Ser-198, and instead has the hydroxyl pointing out toward solvent. Visualization of the density maps supports the close contact between Ser-198 and the phenol hydroxyl (Figure 2B). It was noted that atoms involved in the interaction were ignored during structure refinement due to incompatibility of the short hydrogen bonds with the force field used during refinement.[65]

*Effect of Force Field Choices on Ligand Binding Modes* The positions of the inhibitors in the equilibrated models were first compared with those in the crystal structures (SI Figure 8). It is clear that the inhibitors move further into the active site and assume binding poses with phenol turned slightly outward. The distribution of distances sampled between the ligand phenol and Ser-198 shows that the ligands move further away to relieve the steric clash with both tested force fields, and largely maintain the hydrogen bond except for 1GJD (Figure 3A). The average distances are still within hydrogen bonding range even though there is sampling of unbonded conformations. 1GJD diverges due to rotation of the phenol group, full rotation causes the hydroxyl to point outward toward solvent and the original hydrogen bond is replaced by interaction with the carbonyl oxygen that links the phenol ring to the benzamidine scaffold (Figure 3B). This alternative binding pose is observed with a higher frequency with GAFF2.

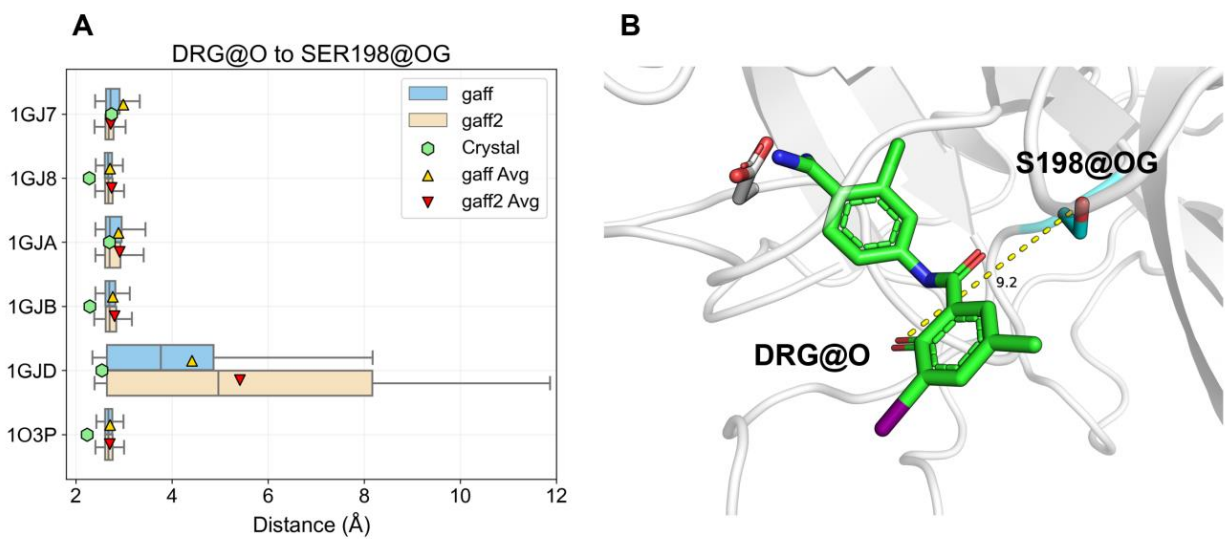

**Figure 3. Relieving steric clash between the ligand phenol and Ser-198.** A) The distance between the ligand phenol oxygen and Ser-198 hydroxyl oxygen is recorded over the last 10 ns of equilibration to analyze sampled conformations and compared to the distance observed in the crystal structures. The trend observed is identical for both GAFF and GAFF2 force fields, all ligands except 1GJD twist away due to repulsive steric interactions but keep in hydrogen bonding range. 1GJD samples broad distances, indicating the initial hydrogen bond is detached. B) Sample frame from the 1GJD simulation illustrates that the phenol hydroxyl rotates outward away from the protein, and the starting hydrogen bond is replaced with one between the peptide bond-like carbonyl and Ser-198. The inhibitor is colored green and labeled DRG.

Next, evaluation of the time evolution of backbone alpha carbon RMSD to crystal, binding pocket RMSD, ligand heavy atom RMSD, and distance from Asp-192 $C_\gamma$ to ligand amidine is performed (SI Figures 9, 10, 11). These values are discretized into 10 ns bins and averaged together from the five replicate trajectories. 1GJD stands out with 0.75 Å binding pocket RMSD using GAFF2 compared to 0.57 Å RMSD with GAFF at the end of equilibration. This major rearrangement is an indication that an alternative binding pose is sampled and agrees with the phenol distance data, showing substantial rotation of the phenol ring. Ligand heavy atom RMSD shows no difference between GAFF and GAFF2. The small set of ligands with fewer torsions cluster together with low RMSD, while the highest RMSD values are observed with 1GJ8 and 1GJD. 1GJD is explained by the phenol rotation. For 1GJ8 the ligand moves away from the crystal pose by sliding more deeply into the binding pocket. The movement into the binding pocket is also observed to a lesser degree with 1O3P, 1GJB, 1GI7, and 1GJA. With both GAFF and GAFF2, the favorable polar interactions between the negatively charged Asp-192 and positively charged amidine draw the ligands into the binding pocket, signaling overestimation of electrostatic interactions that is typical of point charge models. In summary, it is clear that a large discrepancy between the crystal structure and equilibrium binding pose is observed with 1GJD and to a lesser extent with 1GJ8, while the remaining models show close agreement, suggesting that the current force field treatment of 1GJD may not sufficiently characterize the important binding interactions observed in the crystal structure.

Benchmarking the Effects of Simulation Conditions on Predictive Accuracy

We analyzed a range of factors including salt concentration, alternative ligand protonation states, and restraint potential that are known to impact alchemical simulation accuracy. These elements play critical roles in highly charged ligand binding interactions and their effects on predictive accuracy have not been thoroughly characterized in absolute alchemical simulations. Salt concentration plays a role in screening the strength of electrostatic interactions, yet consideration of physiologically relevant salt conditions is often ignored, and counter-ions are generally added only up to the amount necessary to neutralize charge to prevent artifacts arising from periodic boundary conditions. In standard MD simulations, the protonation states of the ligands are fixed and potential changes due to tautomerization or pKa shifts from differences in the solvent and protein environments are not accounted for, which leads to inaccuracy when considering ligands that undergo protonation changes during the binding process. Finally, two types of restraint potentials, 1DOF and 6DOF, have been utilized to prevent the ligand from drifting out of the active site as binding interactions are decoupled, the purpose of these restraints is to focus conformational sampling on configurations most relevant to the binding pose and aid convergence.

*Default Setup Leads to No Correlation with Experiment* The benchmarks begin with a baseline binding free energy prediction to determine the accuracy with a simple and widely accepted model setup, based on a single harmonic distance restraint between the $C_\alpha$ on Asp-192 to the amidine carbon (i.e. 1DOF), with counter-ions added only to the amount to neutralize the system charge, and full ligand protonation without consideration of the experimental data. The Root Mean Square Error (RMSE) for the baseline prediction is 3.2 kcal/mol, and the Pearson correlation coefficient is -0.15, indicating no linear correlation between the experimentally determined binding affinities and those predicted from simulation (Figure 4A). The ligands with 0 net charge form a cluster of samples with underestimated binding free energies, while the charged ligands are predicted to have overestimated free energies indicating excessively favorable binding.

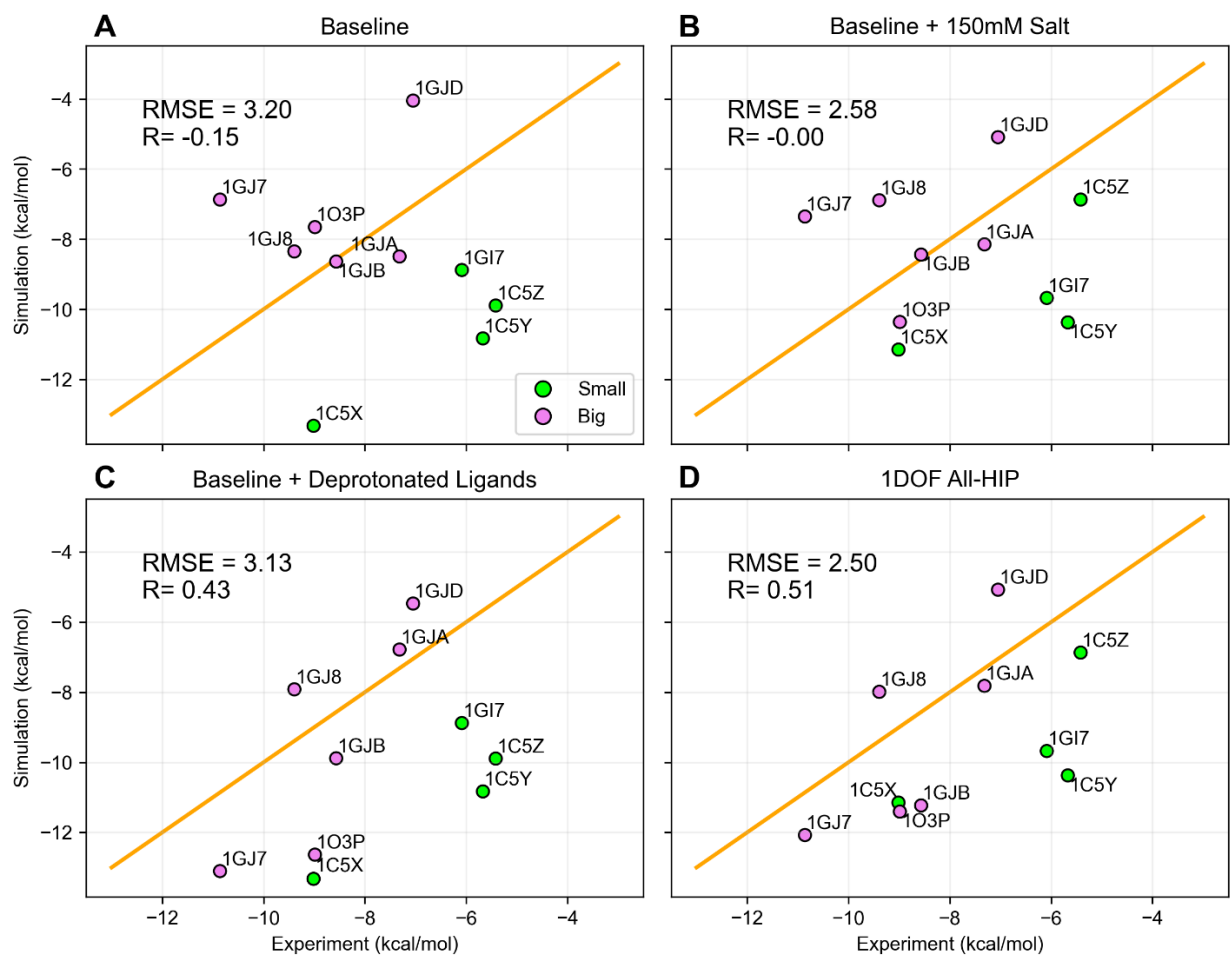

**Figure 4. Baseline absolute alchemical binding predictions for UPA inhibitors.** Evaluating the effects of simulation with 150 mM salt alone, deprotonated ligands alone, and with 150 mM salt and deprotonated ligands combined (1DOF All-HIP) on the baseline condition (fully protonated ligands, counter-ions added only up to neutralize system charge, and 1DOF restraints). The highest performance is observed with the 1DOF All-HIP condition with RMSE 2.50 kcal/mol and Pearson correlation 0.51.

*Use of Salt and Consistent Protonation State Improves Predicted Affinities* Many automated setups neglect setting simulation parameters to match the physiologically relevant conditions, either due to lack of information or to simplify the protocol for higher computation throughput. One often overlooked condition is the salt concentration. The oversight may not be an issue for most neutral or hydrophobic ligands but becomes an important issue for charged ligands due to the impact of ions on electrostatic screening. Given otherwise identical setup and identical restraint as the baseline, the use of 150 mM salt concentration reduced RMSE to 2.58 kcal/mol and improved the Pearson correlation from negative to 0 (Figure 4B). Another discounted issue is the treatment of protonation state for the ligands and amino acid side chains in the binding pocket, which is critical for defining the polar interactions that retain charged ligands. Consistent protonation states maintain hydrogen bond donor and acceptor pairing and/or charge complementary. Modification from the baseline using the deprotonated ligands was found to have minimal improvement in RMSE to 3.13 kcal/mol and significant improvement on the Pearson correlation to 0.43 (Figure 4C). Further, when 150 mM salt concentration and deprotonated ligands are combined,

RMSE is noticeably reduced to 2.50 kcal/mol, and the Pearson correlation is further boosted to 0.51 (Figure 4D). These comparisons highlight a consistent beneficial effect in improving both accuracy and correlation by matching the physiologically relevant salt conditions, and a greatly improved correlation when maintaining consistent protonation states.

*6DOF versus 1DOF in Predicted Affinities* Both 1DOF and 6DOF restraints are widely utilized, but direct comparison has been lacking. The single distance restraint is simpler to implement and enables broader sampling of the binding pocket volume available but has been noted to require longer simulation to reach convergence and may be contaminated by erroneously high energies when the ligand is trapped in a local minimum.[87] The 6DOF approach more tightly locks the ligand into a predefined conformation with limited translational and rotational mobility to more readily achieve convergence and is dependent on securely holding the ligand in the pose that is physically relevant to binding. The binding energy predictions from the 6DOF simulation were found to have an RMSE of 5.59 kcal/mol and Pearson correlation 0.74 (SI Figure 12). The higher Pearson correlation observed with the 6DOF restraints enables a more accurate ranking of the binding energies and may be due to restricting the ligand conformational sampling to a small number of dominant and energetically favorable poses. Indeed, the predicted binding affinities for the 6DOF runs are all more negative than the experimentally determined values, consistent with the ligand being trapped in an excessively favorable binding mode with hindered sampling of higher energy states that are relevant to binding. This reflects how the entropic component is improperly estimated due to the more intensive restriction on sampling. It should also be pointed out that the higher correlation here is likely due to the more negative binding affinities spanning a larger range, indicating use of correlation alone may be insufficient in evaluating the performance.

*Possible Protonation States at Active Sites* We next evaluated the effects of varying the ligand and binding pocket protonation, with deprotonated ligand phenol and Hip-46 or protonated ligand phenol and Hid-46 on predictive accuracy as both satisfy the steric constraint in the crystal structures. Assignment of hydrogens is typically not resolved with structure determination by X-ray crystallography. The issue is further complicated by the absence of direct pKa measurement for the system. Nevertheless, based on the close distance between the ligand phenol hydroxyl and Ser-198, Katz et al. inferred that the ligand phenol binds as an acid and is deprotonated to minimize steric clash with surrounding atoms.[65] The free oxygen then acts as a hydrogen bond acceptor interacting with Hip-46. However, the typical pKa of a phenol hydroxyl is approximately 10 and those on the ligands range between 8-9,[65, 118] which suggests that maintenance of the hydroxyl proton is favored under physiological conditions. His-46 would more likely assume the neutral HID form allowing hydrogen bonding to occur at $N_\varepsilon$ on Hid-46. To investigate both possibilities, trials were conducted in four groups as all-HID (all ligands interacting with Hid-46), all-HIP (all ligands interacting with Hip-46), small-HID (larger phenol ligands interacting with Hip-46 and smaller non-phenol ligands interacting with Hid-46), and small-HIP (larger phenol ligands interacting with Hid-46 and smaller non-phenol ligands interacting with Hip-46).

   Utilization of deprotonated ligands and all Hip-46 (all-HIP) led to RMSE of 2.50 kcal/mol and Pearson correlation of 0.51 as previously shown in Figure 4. In contrast, the alternative with protonated ligands and all Hid-46 (all-HID) resulted in a worse RMSE of 3.91 kcal/mol and slightly reduced Pearson correlation of 0.47 (SI Figure 13). Since the smaller and non-phenol ligands are not expected to form hydrogen-bonding contact with His-46, the protonation state of His-46 may not match that for the larger ligands. Thus, a more appropriate comparison is between small-HID versus all-HID. Interestingly the small-HID condition leads to RMSE 3.40 kcal/mol and Pearson correlation 0.21 (SI Figure 13). For the fourth small-HIP condition, RMSE was calculated to be 3.16 kcal/mol with the highest Pearson correlation of 0.69 (SI Figure 13).

Notably, 1GJD is an outlier in all conditions, separated from the cluster of other ligands and is predicted to have higher binding free energy than measured in experiment for both HIP and HID conditions. This is potentially due to force field imperfections as discussed above: the 1GJD phenol pivots away from Ser-198 observed during the equilibration phase (Figure 3B). All other ligands adopted poses with the phenol shifted away from Ser-198 slightly to alleviate steric clash but maintained hydrogen bonding range. Thus, 1GJD is excluded from further binding analysis. Removal of 1GJD from aggregate calculations does not improve RMSE as it increased to 2.55, 4.06, 3.52, and 3.25 kcal/mol for the all-HIP, all-HID, small-HID, and small-HIP conditions, respectively, but its omission increases Pearson correlations for all-HIP to 0.55, all-HID to 0.81, reduced small-HID to 0.14, and brings small-HIP to 0.85 (SI Figures 13). These simulations demonstrate the impact of protonation state on the binding free energy prediction. Our standard alchemical simulations suggest that the all-HIP condition with lowest RMSE and all-HID and small-HIP conditions with over 80% correlation may all explain some aspects of the experimental binding affinities. However, the absolute errors are all quite large, over 2.5 kcal/mol which is above the chemically accurate threshold of 1.0 kcal/mol. Therefore, it is still uncertain which binding mode best describes these challenging systems.

Estimation of Electronic Polarization by MBAR/PBSA

Following optimization of protein and ligand radii for all alchemical conditions tested (SI Table 1), we evaluated the effect of solute dielectric on the accuracy of binding affinities to assess the impact of incorporating polarization into the computational models. Evaluation of solute dielectric at the theoretical value of 2 responsible for electronic polarization[119-122] shows improved Pearson correlation to as high as 0.81, highlighting its applicability to correctly ranking candidate inhibitors by offsetting charge polarization errors. However, the RMSE increases to as high as 4.84 kcal/mol (SI Table 2). All samples are predicted to have more positive binding free energies with the increasing solute dielectrics, demonstrating that screening charged effects increase the predicted free energies.

The standard Amber force fields were developed with effective partial charges to model electrostatics and include polarization responses to the environment (mostly in water), though only in an averaged, mean-field manner. They are not fully compatible with the theoretical dielectric constant of 2 because polarization is already partially accounted for in the effective partial charges. Thus, a further scanning procedure to find the optimal solute dielectrics is necessary. In doing so, the RMSE to experimental affinities was found to be reduced to as low as 0.89 kcal/mol and the Pearson correlation is increased to as high as 0.88 for the all-HID condition (Figure 5, SI Table 2). Both metrics are dramatically improved compared to the explicit solvent simulation, enabling more accurate binding free energy prediction with only post-processing of existing trajectory data and minimal modification to current protocols.

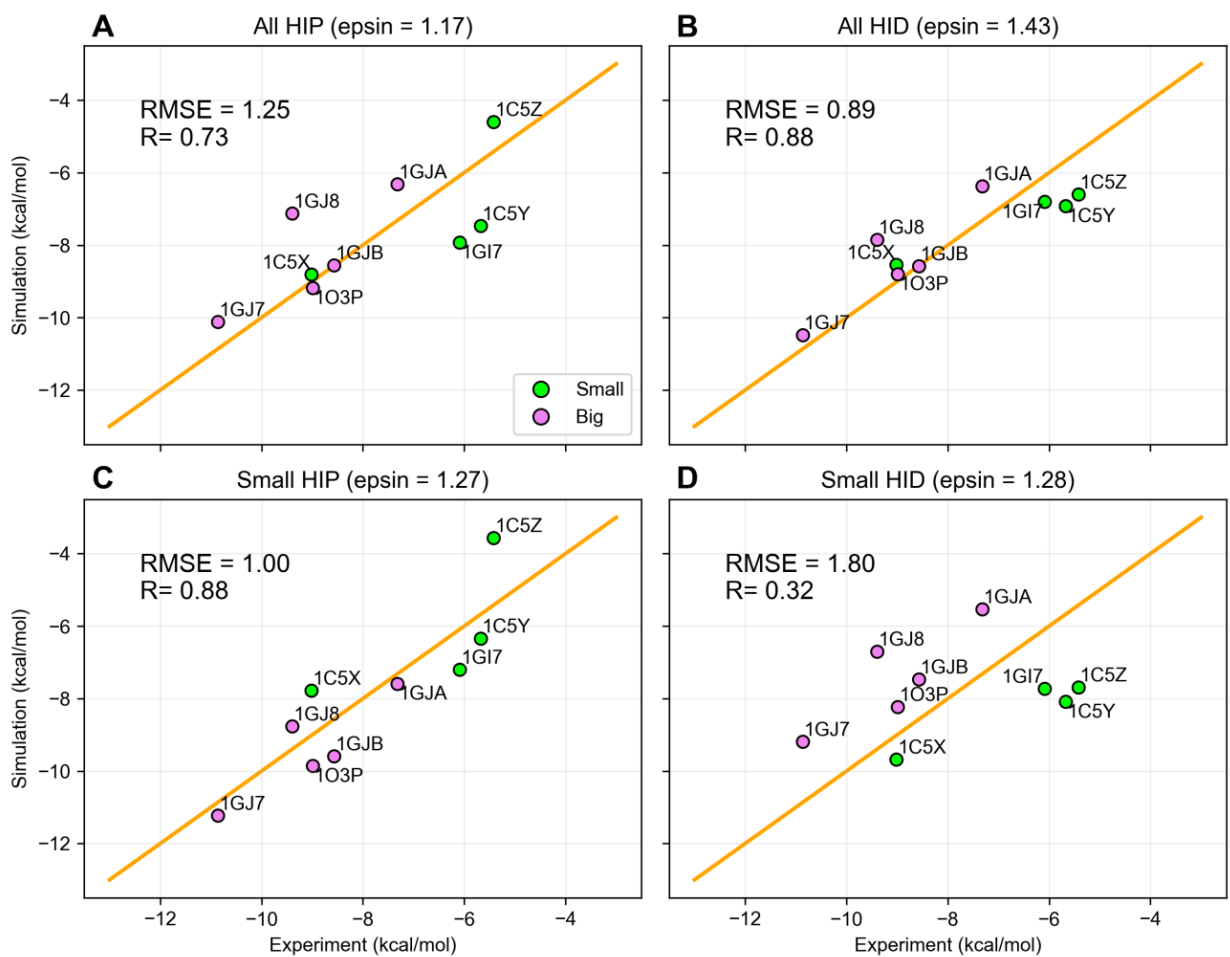

**Figure 5. MBAR/PBSA binding affinity calculations.** The All-HID condition shows the best agreement to experiment with consideration of polarization effects through solute dielectric scaling. In comparison to values from the standard alchemical transformation, RMSE's are reduced and Pearson correlations are improved for all conditions.

It is interesting to see how accounting for polarization affects the prediction of binding free energy at the tested alternative protonation states. Among the three viable candidates from standard alchemical simulations, all-HIP, all-HID, and small-HIP conditions, the all-HIP condition is improved to 1.25 kcal/mol RMSE and 0.74 Pearson correlation, the all-HID condition is calculated to have 0.89 kcal/mol RMSE and 0.88 Pearson correlation, and the small-HIP condition is changed to 1.0 kcal/mol RMSE and 0.88 Pearson correlation. The last tested condition, small-HID, has 1.80 kcal/mol RMSE and 0.32 Pearson correlation.

This suggests that the environment of the binding pocket and charged nature of the ligands may not fully support the originally proposed binding mode with deprotonated phenol and Hip-46. Instead, the alternative hypothesis of the ligands with protonated phenol as a hydrogen bond donor and Hid-46 assuming the role of hydrogen bond acceptor may better describe the protein-ligand interactions. Both protonation states satisfy the steric constraint in the crystal structures. However, the differences in errors are within kT between different protonation states. Our analysis here points to the need for more definite NMR pKa measurement to resolve the issue.[123, 124]

**Conclusion**

The current study aims to understand the impact of simulation conditions for absolute binding calculations in the UPA system, introduces the MBAR/PBSA continuum solvent approach in the calculation of decharging free energies to capture electronic polarization effects absent in standard explicit solvent models, and evaluates the effect of varying protonation states of titratable ligands and protein residues in binding free energy prediction. Extensive simulations of UPA with a broad set of inhibitors were performed to benchmark the performance of absolute alchemical simulations, which have been sparsely studied due to their demanding calculation, allowing us to identify factors pivotal to increasing predictive accuracy. The force field description of the ligands plays a significant role in maintaining important interactions and poses for binding, and issues with the ligand force field parameters can cause inaccuracies in the binding calculation as seen in the case of 1GJD. Here, difficulty maintaining the short hydrogen bond between the ligand phenol and Ser-198 caused excessive rotation of the phenol ring and also overly positive binding energy prediction. Furthermore, the setup of simulation systems contributes significantly to predictive accuracy as seen in the baseline condition, which does not account for salt concentration, protonation state, or polarization effects. These oversights lead to poor performance in prediction with 3.2 kcal/mol RMSE and -0.15 Pearson correlation. As the simulation conditions are modified to be consistent with physiologically relevant conditions, notable improvements in the accuracy are observed with the RMSE decreasing to 2.5 kcal/mol and the increase in Pearson correlation to 0.51. The more restrictive 6DOF ligand restraints were found to overestimate binding affinity by keeping the ligand in a singular binding conformation, preventing exploration of relevant higher energy conformations and resulting in larger error, but improved Pearson correlation compared to 1DOF restraints.

| Condition | Method | RMSE (kcal/mol) | R |
|---|---|---|---|
| Baseline | Standard alchemical | 3.22 (3.20) | -0.34 (-0.15) |
| Baseline + 150 mM salt | Standard alchemical | 2.64 (2.58) | -0.12 (0.00) |
| Baseline + Deprotonated Ligands | Standard alchemical | 3.25 (3.13) | 0.42 (0.43) |
| 6DOF (All-HIP) | Standard alchemical | 5.85 (5.59) | 0.75 (0.74) |
| 1DOF (All-HIP) | Standard alchemical | 2.55 (2.50) | 0.55 (0.51) |
| All-HID | Standard alchemical | 4.06 (3.91) | 0.81 (0.47) |
| Small-HIP | Standard alchemical | 3.25 (3.16) | 0.85 (0.69) |
| Small-HID | Standard alchemical | 3.52 (3.40) | 0.14 (0.21) |
| All-HIP | MBAR/PBSA | 1.25 (1.61) | 0.73 (0.65) |
| All-HID | MBAR/PBSA | 0.89 (1.53) | 0.88 (0.67) |
| Small-HIP | MBAR/PBSA | 1.00 (1.48) | 0.88 (0.81) |
| Small-HID | MBAR/PBSA | 1.80 (2.18) | 0.32 (0.31) |

**Table 2. Summary of error and correlation statistics.** Binding free energy prediction metrics with outlier 1GJD removed. Values in parenthesis represent inclusion of the outlier. Conditions examining binding pocket protonation include the simulation with 150 mM salt and deprotonated ligands (1DOF All-HIP). The Baseline condition is described by inclusion of only neutralizing counter-ions and with fully protonated ligand phenol groups.

Importantly, simulation conditions that affect electrostatic interactions are observed to have a major contribution to binding prediction accuracy, augmenting results from previous studies.[57, 125] Standard MD simulation utilizing explicit solvation and point-charge models lack the capability to account for electronic polarization effects that undoubtedly occur as the ligands transition from the high-dielectric water environment to the low-dielectric protein interior. Polarization effects can be captured through ab initio quantum calculations that evaluate the electron densities surrounding each atom, but their usage is limited by steep computational costs and typically require that the system is separated into coupled QM/MM regions where the choice of boundary and level of QM theory entangle accurate treatment.[126, 127] The MBAR/PBSA calculation allows the assignment of different dielectric values to solute and solvent, enabling us to measure the impact of including polarization effects on binding affinity prediction. The interior dielectric constant in PBSA parameterizes the strength of charge screening in the protein environment. At the default value 1, atom charges are not shielded resulting in exaggerated attractive and repulsive interactions as the atom partial charges, typically assigned for the ligand in gas phase, cannot be adjusted in standard MD simulation. This overestimation of the electrostatic potential can be offset by finely increasing the solute dielectric value to imitate the effect of electronic polarization that masks electrostatics. When the active site protonation state is defined and the polarization effects are modeled in the MBAR/PBSA calculation, significant enhancement of prediction accuracy is observed. This method is a mean-field approach demonstrated here for its ease of implementation and inspires the utilization of more explicit calculations of electronic polarization such as with polarizable multipole electrostatics.[9, 13, 128]

The all-HIP condition was first inferred to be a likely protonation state satisfies the crystal steric constraint. It was found to have 1.25 kcal/mol RMSE and 0.73 Pearson correlation, but is not necessarily the definitive state as the alternative protonation state, all-HID, shows higher prediction accuracy, though the differences in errors are within kT. Conclusive protonation assignment requires further experimental validation such as pKa determination via NMR spectroscopy.[123, 124] This is significant when considering how simulation protocols and algorithms deal with aspects of electrostatic interactions in defining protonation states and handling polarization effects.

Complete examination of protonation changes is limited with existing simulation protocols. Exploring alternative protonation states becomes an important and complicating process if the proton transfer events of the whole system are coupled to the binding process. In the Supporting Information, the calculation of the contributions of a single titratable group coupled to the binding process is discussed. However, this simple model is inadequate for most protein systems, as they often have multiple titratable groups that are coupled directly or through long- range allosteric interactions.[129, 130] In particular, the current UPA system involves titratable residues in the active site at His-46, His-94, and Asp-97, and titratable functional groups on several ligands such as the phenol hydroxyl. Direct interaction may shift the pKa of the titratable groups involved. The investigation of how to approach these coupled processes has been explored by several groups[131-134] and includes approaches ranging from corrections as discussed in the SI,[132, 133] the explicit enumeration of protonation states for the binding simulations, and techniques such as constant pH molecular dynamics.[134]

**Supporting Information**

Discussion of apparent binding free energies with one titratable group, MBAR/PBSA accuracy, binding free energy predictions for all conditions, illustration of restraint definitions, decharging and VDW free energies from alchemical simulations, illustration of the UPA active site with ligand, B-Factor, GAFF and GAFF2 structure analysis, and correlation plots of binding affinity for differing protonation states, with and without outlier and with MBAR/PBSA postprocessing.

## Acknowledgements

This work was supported by National Institute of Health/NIGMS (Grant Nos. GM093040 and GM130367)

# References

1. Clementi, E.; Kistenmacher, H.; Kolos, W.; Romano, S., Non-Additivity in Water-Ion-Water Interactions. *Theor. Chim. Acta* **1980,** *55* (4), 257-266.
2. Ren, P. Y.; Ponder, J. W., Polarizable atomic multipole water model for molecular mechanics simulation. *J. Phys. Chem. B* **2003,** *107* (24), 5933-5947.
3. Kaminski, G. A.; Stern, H. A.; Berne, B. J.; Friesner, R. A.; Cao, Y. X. X.; Murphy, R. B.; Zhou, R. H.; Halgren, T. A., Development of a polarizable force field for proteins via ab initio quantum chemistry: First generation model and gas phase tests. *J. Comput. Chem.* **2002,** *23* (16), 1515-1531.
4. Friesner, R. A., Modeling Polarization in Proteins and Protein-Ligand Complexes: Methods and Preliminary Results. *Adv. Protein Chem.* **2005,** *72*, 79-104.
5. Patel, S.; Mackerell, A. D.; Brooks, C. L., CHARMM fluctuating charge force field for proteins: II - Protein/solvent properties from molecular dynamics simulations using a nonadditive electrostatic model. *J. Comput. Chem.* **2004,** *25* (12), 1504-1514.
6. Lopes, P. E. M.; Lamoureux, G.; Roux, B.; MacKerell, A. D., Polarizable empirical force field for aromatic compounds based on the classical drude oscillator. *J. Phys. Chem. B* **2007,** *111* (11), 2873-2885.
7. Jiang, W.; Hardy, D. J.; Phillips, J. C.; MacKerell, A. D.; Schulten, K.; Roux, B., High-Performance Scalable Molecular Dynamics Simulations of a Polarizable Force Field Based on Classical Drude Oscillators in NAMD. *J. Phys. Chem. Lett.* **2011,** *2* (2), 87-92.
8. Lamoureux, G.; Harder, E.; Vorobyov, I. V.; Roux, B.; MacKerell, A. D., A polarizable model of water for molecular dynamics simulations of biomolecules. *Chem. Phys. Lett.* **2006,** *418* (1-3), 245-249.
9. Ponder, J. W.; Wu, C. J.; Ren, P. Y.; Pande, V. S.; Chodera, J. D.; Schnieders, M. J.; Haque, I.; Mobley, D. L.; Lambrecht, D. S.; DiStasio, R. A.; Head-Gordon, M.; Clark, G. N. I.; Johnson, M. E.; Head-Gordon, T., Current Status of the AMOEBA Polarizable Force Field. *J. Phys. Chem. B* **2010,** *114* (8), 2549-2564.
10. Shi, Y.; Xia, Z.; Zhang, J. J.; Best, R.; Wu, C. J.; Ponder, J. W.; Ren, P. Y., Polarizable Atomic Multipole-Based AMOEBA Force Field for Proteins. *J. Chem. Theory Comput.* **2013,** *9* (9), 4046-4063.
11. Zhang, C. S.; Lu, C.; Jing, Z. F.; Wu, C. J.; Piquemal, J. P.; Ponder, J. W.; Ren, P. Y., AMOEBA Polarizable Atomic Multipole Force Field for Nucleic Acids. *J. Chem. Theory Comput.* **2018,** *14* (4), 2084-2108.
12. Wang, J.; Cieplak, P.; Luo, R.; Duan, Y., Development of Polarizable Gaussian Model for Molecular Mechanical Calculations I: Atomic Polarizability Parameterization To Reproduce ab Initio Anisotropy. *J. Chem. Theory Comput.* **2019,** *15* (2), 1146-1158.
13. Wei, H. X.; Qi, R. X.; Wang, J. M.; Cieplak, P.; Duan, Y.; Luo, R., Efficient formulation of polarizable Gaussian multipole electrostatics for biomolecular simulations. *J. Chem. Phys.* **2020,** *153* (11).
14. Cieplak, P.; Caldwell, J.; Kollman, P., Molecular mechanical models for organic and biological systems going beyond the atom centered two body additive approximation: Aqueous solution free energies of methanol and N-methyl acetamide, nucleic acid base, and amide hydrogen bonding and chloroform/water partition coefficients of the nucleic acid bases. *J. Comput. Chem.* **2001,** *22* (10), 1048-1057.
15. Wang, J. M.; Cieplak, P.; Li, J.; Hou, T. J.; Luo, R.; Duan, Y., Development of Polarizable Models for Molecular Mechanical Calculations I: Parameterization of Atomic Polarizability. *J. Phys. Chem. B* **2011,** *115* (12), 3091-3099.
16. Wang, J. M.; Cieplak, P.; Li, J.; Wang, J.; Cai, Q.; Hsieh, M. J.; Lei, H. X.; Luo, R.; Duan, Y., Development of Polarizable Models for Molecular Mechanical Calculations II: Induced Dipole

Models Significantly Improve Accuracy of Intermolecular Interaction Energies. *J. Phys. Chem. B* **2011,** *115* (12), 3100-3111.
17. Wang, J.; Cieplak, P.; Cai, Q.; Hsieh, M. J.; Wang, J. M.; Duan, Y.; Luo, R., Development of Polarizable Models for Molecular Mechanical Calculations. 3. Polarizable Water Models Conforming to Thole Polarization Screening Schemes. *J. Phys. Chem. B* **2012,** *116* (28), 7999-8008.
18. Wang, J. M.; Cieplak, P.; Li, J.; Cai, Q.; Hsieh, M. J.; Luo, R.; Duan, Y., Development of Polarizable Models for Molecular Mechanical Calculations. 4. van der Waals Parametrization. *J. Phys. Chem. B* **2012,** *116* (24), 7088-7101.
19. Gilson, M. K.; Zhou, H.-X., Calculation of Protein-Ligand Binding Affinities. *Annu. Rev. Biophys. Biomol. Struct.* **2007,** *36* (1), 21-42.
20. DiMasi, J. A.; Grabowski, H. G.; Hansen, R. W., Innovation in the pharmaceutical industry: New estimates of R&D costs. *J. Health Econ.* **2016,** *47*, 20-33.
21. DeLuca, S.; Khar, K.; Meiler, J., Fully Flexible Docking of Medium Sized Ligand Libraries with RosettaLigand. *PLoS One* **2015,** *10* (7), e0132508.
22. Clark, A. J.; Tiwary, P.; Borrelli, K.; Feng, S.; Miller, E. B.; Abel, R.; Friesner, R. A.; Berne, B. J., Prediction of Protein-Ligand Binding Poses via a Combination of Induced Fit Docking and Metadynamics Simulations. *J. Chem. Theory Comput.* **2016,** *12* (6), 2990-8.
23. Olsson, M. A.; Garcia-Sosa, A. T.; Ryde, U., Binding affinities of the farnesoid X receptor in the D3R Grand Challenge 2 estimated by free-energy perturbation and docking. *J. Comput. Aided Mol. Des.* **2018,** *32* (1), 211-224.
24. Lyu, J.; Wang, S.; Balius, T. E.; Singh, I.; Levit, A.; Moroz, Y. S.; O'Meara, M. J.; Che, T.; Algaa, E.; Tolmachova, K.; Tolmachev, A. A.; Shoichet, B. K.; Roth, B. L.; Irwin, J. J., Ultra-large library docking for discovering new chemotypes. *Nature* **2019,** *566* (7743), 224-229.
25. David, L.; Luo, R.; Gilson, M. K., Ligand-receptor docking with the Mining Minima optimizer. *J. Comput. Aided Mol. Des.* **2001,** *15* (2), 157-171.
26. Åqvist, J.; Medina, C.; Samuelsson, J.-E., A new method for predicting binding affinity in computer-aided drug design. *Protein Eng. Des. Sel.* **1994,** *7* (3), 385-391.
27. Miller III, B. R.; McGee Jr, T. D.; Swails, J. M.; Homeyer, N.; Gohlke, H.; Roitberg, A. E., MMPBSA.py: an efficient program for end-state free energy calculations. *J. Chem. Theory Comput.* **2012,** *8* (9), 3314-3321.
28. Wang, J.; Cai, Q.; Xiang, Y.; Luo, R., Reducing Grid Dependence in Finite-Difference Poisson-Boltzmann Calculations. *J. Chem. Theory Comput.* **2012,** *8* (8), 2741-2751.
29. Wang, C.; Nguyen, P. H.; Pham, K.; Huynh, D.; Le, T. B.; Wang, H.; Ren, P.; Luo, R., Calculating protein-ligand binding affinities with MMPBSA: Method and error analysis. *J. Comput. Chem.* **2016,** *37* (27), 2436-46.
30. Xiao, L.; Diao, J.; Greene, D. A.; Wang, J.; Luo, R., A Continuum Poisson–Boltzmann Model for Membrane Channel Proteins. *J. Chem. Theory Comput.* **2017,** *13* (7), 3398-3412.
31. Qi, R.; Botello-Smith, W. M.; Luo, R., Acceleration of Linear Finite-Difference Poisson–Boltzmann Methods on Graphics Processing Units. *J. Chem. Theory Comput.* **2017,** *13* (7), 3378-3387.
32. Wang, C.; Greene, D. A.; Xiao, L.; Qi, R.; Luo, R., Recent Developments and Applications of the MMPBSA Method. *Front. Mol. Biosci.* **2018,** *4* (87).
33. Greene, D.; Qi, R.; Nguyen, R.; Qiu, T.; Luo, R., Heterogeneous Dielectric Implicit Membrane Model for the Calculation of MMPBSA Binding Free Energies. *J. Chem. Inf. Model.* **2019,** *59* (6), 3041-3056.
34. Qi, R.; Luo, R., Robustness and Efficiency of Poisson–Boltzmann Modeling on Graphics Processing Units. *J. Chem. Inf. Model.* **2019,** *59* (1), 409-420.

35. Wei, H.; Luo, R.; Qi, R., An efficient second-order poisson–boltzmann method. *J. Comput. Chem.* **2019,** *40* (12), 1257-1269.
36. Wei, H.; Luo, A.; Qiu, T.; Luo, R.; Qi, R., Improved Poisson–Boltzmann Methods for High-Performance Computing. *J. Chem. Theory Comput.* **2019,** *15* (11), 6190-6202.
37. Gilson, M. K.; Given, J. A.; Bush, B. L.; McCammon, J. A., The Statistical-Thermodynamic Basis for Computation of Binding Affinities: A Critical Review. *Biophys. J.* **1997,** *72*, 1047.
38. Luo, R.; Head, M. S.; Moult, J.; Gilson, M. K., pK(a) shifts in small molecules and HIV protease: Electrostatics and conformation. *J. Am. Chem. Soc.* **1998,** *120* (24), 6138-6146.
39. Luo, R.; Head, M. S.; Given, J. A.; Gilson, M. K., Nucleic acid base-pairing and N-methylacetamide self-association in chloroform: affinity and conformation. *Biophys. Chem.* **1999,** *78* (1-2), 183-193.
40. Chang, C.-e. A.; Chen, W.; Gilson, M. K., Ligand configurational entropy and protein binding. *Proc. Natl. Acad. Sci.* **2007,** *104* (5), 1534-1539.
41. Zou, J.; Tian, C.; Simmerling, C., Blinded prediction of protein-ligand binding affinity using Amber thermodynamic integration for the 2018 D3R grand challenge 4. *J. Comput. Aided Mol. Des.* **2019,** *33* (12), 1021-1029.
42. Kim, M. O.; Blachly, P. G.; McCammon, J. A., Conformational Dynamics and Binding Free Energies of Inhibitors of BACE-1: From the Perspective of Protonation Equilibria. *PLoS Comput. Biol.* **2015,** *11* (10), e1004341.
43. Giese, T. J.; York, D. M., A GPU-Accelerated Parameter Interpolation Thermodynamic Integration Free Energy Method. *J. Chem. Theory Comput.* **2018,** *14* (3), 1564-1582.
44. Lee, T. S.; Hu, Y.; Sherborne, B.; Guo, Z.; York, D. M., Toward Fast and Accurate Binding Affinity Prediction with pmemdGTI: An Efficient Implementation of GPU-Accelerated Thermodynamic Integration. *J. Chem. Theory Comput.* **2017,** *13* (7), 3077-3084.
45. Aldeghi, M.; Heifetz, A.; Bodkin, M. J.; Knapp, S.; Biggin, P. C., Predictions of Ligand Selectivity from Absolute Binding Free Energy Calculations. *J. Am. Chem. Soc.* **2017,** *139* (2), 946-957.
46. Lapelosa, M.; Gallicchio, E.; Levy, R. M., Conformational Transitions and Convergence of Absolute Binding Free Energy Calculations. *J. Chem. Theory Comput.* **2012,** *8* (1), 47-60.
47. Jiang, W.; Roux, B., Free Energy Perturbation Hamiltonian Replica-Exchange Molecular Dynamics (FEP/H-REMD) for Absolute Ligand Binding Free Energy Calculations. *J. Chem. Theory Comput.* **2010,** *6* (9), 2559-2565.
48. Ragoza, M.; Hochuli, J.; Idrobo, E.; Sunseri, J.; Koes, D. R., Protein-Ligand Scoring with Convolutional Neural Networks. *J. Chem. Inf. Model.* **2017,** *57* (4), 942-957.
49. Ericksen, S. S.; Wu, H.; Zhang, H.; Michael, L. A.; Newton, M. A.; Hoffmann, F. M.; Wildman, S. A., Machine Learning Consensus Scoring Improves Performance Across Targets in Structure-Based Virtual Screening. *J. Chem. Inf. Model.* **2017,** *57* (7), 1579-1590.
50. Pereira, J. C.; Caffarena, E. R.; Dos Santos, C. N., Boosting Docking-Based Virtual Screening with Deep Learning. *J. Chem. Inf. Model.* **2016,** *56* (12), 2495-2506.
51. vKlimovich, P. V.; Shirts, M. R.; Mobley, D. L., Guidelines for the analysis of free energy calculations. *J. Comput. Aided Mol. Des.* **2015,** *29* (5), 397-411.
52. Shirts, M. R.; Mobley, D. L.; Chodera, J. D., Chapter 4 Alchemical Free Energy Calculations: Ready for Prime Time? 2007; pp 41-59.
53. Rustenburg, A. S.; Dancer, J.; Lin, B.; Feng, J. A.; Ortwine, D. F.; Mobley, D. L.; Chodera, J. D., Measuring experimental cyclohexane-water distribution coefficients for the SAMPL5 challenge. *J. Comput. Aided Mol. Des.* **2016,** *30* (11), 945-958.
54. Gapsys, V.; Michielssens, S.; Seeliger, D.; de Groot, B. L., Accurate and Rigorous Prediction of the Changes in Protein Free Energies in a Large-Scale Mutation Scan. *Angew. Chem., Int. Ed. Engl.* **2016,** *55* (26), 7364-8.

55. de Oliveira, C.; Yu, H. S.; Chen, W.; Abel, R.; Wang, L., Rigorous Free Energy Perturbation Approach to Estimating Relative Binding Affinities between Ligands with Multiple Protonation and Tautomeric States. *J. Chem. Theory Comput.* **2019,** *15* (1), 424-435.
56. Cournia, Z.; Allen, B.; Sherman, W., Relative Binding Free Energy Calculations in Drug Discovery: Recent Advances and Practical Considerations. *J. Chem. Inf. Model.* **2017,** *57* (12), 2911-2937.
57. Chen, W.; Deng, Y.; Russell, E.; Wu, Y.; Abel, R.; Wang, L., Accurate Calculation of Relative Binding Free Energies between Ligands with Different Net Charges. *J. Chem. Theory Comput.* **2018,** *14* (12), 6346-6358.
58. Li, Z.; Huang, Y.; Wu, Y.; Chen, J.; Wu, D.; Zhan, C. G.; Luo, H. B., Absolute Binding Free Energy Calculation and Design of a Subnanomolar Inhibitor of Phosphodiesterase-10. *J. Med. Chem.* **2019,** *62* (4), 2099-2111.
59. Aldeghi, M.; Heifetz, A.; Bodkin, M. J.; Knapp, S.; Biggin, P. C., Accurate calculation of the absolute free energy of binding for drug molecules. *Chem. Sci.* **2016,** *7* (1), 207-218.
60. Qian, Y.; Cabeza de Vaca, I.; Vilseck, J. Z.; Cole, D. J.; Tirado-Rives, J.; Jorgensen, W. L., Absolute Free Energy of Binding Calculations for Macrophage Migration Inhibitory Factor in Complex with a Druglike Inhibitor. *J. Phys. Chem. B* **2019,** *123* (41), 8675-8685.
61. Okamoto, Y.; Kokubo, H.; Tanaka, T., Prediction of Ligand Binding Affinity by the Combination of Replica-Exchange Method and Double-Decoupling Method. *J. Chem. Theory Comput.* **2014,** *10* (8), 3563-9.
62. Deng, N.; Wickstrom, L.; Cieplak, P.; Lin, C.; Yang, D., Resolving the Ligand-Binding Specificity in c-MYC G-Quadruplex DNA: Absolute Binding Free Energy Calculations and SPR Experiment. *J. Phys. Chem. B* **2017,** *121* (46), 10484-10497.
63. Mahmood, N.; Mihalcioiu, C.; Rabbani, S. A., Multifaceted Role of the Urokinase-Type Plasminogen Activator (uPA) and Its Receptor (uPAR): Diagnostic, Prognostic, and Therapeutic Applications. *Front. Oncol.* **2018,** *8*, 24.
64. Katz, B. A.; Elrod, K.; Luong, C.; Rice, M. J.; Mackman, R. L.; Sprengeler, P. A.; Spencer, J.; Hataye, J.; Janc, J.; Link, J.; Litvak, J.; Rai, R.; Rice, K.; Sideris, S.; Verner, E.; Young, W., A novel serine protease inhibition motif involving a multi-centered short hydrogen bonding network at the active site. *J. Mol. Biol.* **2001,** *307* (5), 1451-86.
65. Katz, B. A.; Elrod, K.; Verner, E.; Mackman, R. L.; Luong, C.; Shrader, W. D.; Sendzik, M.; Spencer, J. R.; Sprengeler, P. A.; Kolesnikov, A.; Tai, V. W.; Hui, H. C.; Breitenbucher, J. G.; Allen, D.; Janc, J. W., Elaborate manifold of short hydrogen bond arrays mediating binding of active site-directed serine protease inhibitors. *J. Mol. Biol.* **2003,** *329* (1), 93-120.
66. Katz, B. A.; Mackman, R.; Luong, C.; Radika, K.; Martelli, A.; Sprengeler, P. A.; Wang, J.; Chan, H.; Wong, L., Structural basis for selectivity of a small molecule, S1-binding, submicromolar inhibitor of urokinase-type plasminogen activator. *Chem. Biol.* **2000,** *7* (4), 299-312.
67. Berman, H. M.; Westbrook, J.; Feng, Z.; Gilliland, G.; Bhat, T. N.; Weissig, H.; Shindyalov, I. N.; Bourne, P. E., The Protein Data Bank. *Nucleic Acids Res.* **2000,** *28* (1), 235-42.
68. Wang, R.; Fang, X.; Lu, Y.; Yang, C. Y.; Wang, S., The PDBbind database: methodologies and updates. *J. Med. Chem.* **2005,** *48* (12), 4111-9.
69. Anandakrishnan, R.; Aguilar, B.; Onufriev, A. V., H++ 3.0: automating pK prediction and the preparation of biomolecular structures for atomistic molecular modeling and simulations. *Nucleic Acids Res.* **2012,** *40* (Web Server issue), W537-41.
70. Bayly, C. I.; Cieplak, P.; Cornell, W. D.; Kollman, P. A., A Well-Behaved Electrostatic Potential Based Method Using Charge Restraints for Deriving Atomic Charges - the Resp Model. *J. Phys. Chem.* **1993,** *97* (40), 10269-10280.
71. Frisch, M. J.; Trucks, G. W.; Schlegel, H. B.; Scuseria, G. E.; Robb, M. A.; Cheeseman, J. R.; Scalmani, G.; Barone, V.; Petersson, G. A.; Nakatsuji, H.; Li, X.; Caricato, M.; Marenich, A. V.;

Bloino, J.; Janesko, B. G.; Gomperts, R.; Mennucci, B.; Hratchian, H. P.; Ortiz, J. V.; Izmaylov, A. F.; Sonnenberg, J. L.; Williams; Ding, F.; Lipparini, F.; Egidi, F.; Goings, J.; Peng, B.; Petrone, A.; Henderson, T.; Ranasinghe, D.; Zakrzewski, V. G.; Gao, J.; Rega, N.; Zheng, G.; Liang, W.; Hada, M.; Ehara, M.; Toyota, K.; Fukuda, R.; Hasegawa, J.; Ishida, M.; Nakajima, T.; Honda, Y.; Kitao, O.; Nakai, H.; Vreven, T.; Throssell, K.; Montgomery Jr., J. A.; Peralta, J. E.; Ogliaro, F.; Bearpark, M. J.; Heyd, J. J.; Brothers, E. N.; Kudin, K. N.; Staroverov, V. N.; Keith, T. A.; Kobayashi, R.; Normand, J.; Raghavachari, K.; Rendell, A. P.; Burant, J. C.; Iyengar, S. S.; Tomasi, J.; Cossi, M.; Millam, J. M.; Klene, M.; Adamo, C.; Cammi, R.; Ochterski, J. W.; Martin, R. L.; Morokuma, K.; Farkas, O.; Foresman, J. B.; Fox, D. J. *Gaussian 16 Rev. C.01*, Wallingford, CT, 2016.
72. Wang, J.; Wolf, R. M.; Caldwell, J. W.; Kollman, P. A.; Case, D. A., Development and testing of a general amber force field. *J. Comput. Chem.* **2004,** *25* (9), 1157-74.
73. Maier, J. A.; Martinez, C.; Kasavajhala, K.; Wickstrom, L.; Hauser, K. E.; Simmerling, C., ff14SB: Improving the Accuracy of Protein Side Chain and Backbone Parameters from ff99SB. *J. Chem. Theory Comput.* **2015,** *11* (8), 3696-3713.
74. Jorgensen, W. L.; Chandrasekhar, J.; Madura, J. D.; Impey, R. W.; Klein, M. L., Comparison of Simple Potential Functions for Simulating Liquid Water. *J. Chem. Phys.* **1983,** *79* (2), 926-935.
75. Salomon-Ferrer, R.; Gotz, A. W.; Poole, D.; Le Grand, S.; Walker, R. C., Routine Microsecond Molecular Dynamics Simulations with AMBER on GPUs. 2. Explicit Solvent Particle Mesh Ewald. *J. Chem. Theory Comput.* **2013,** *9* (9), 3878-3888.
76. Essmann, U.; Perera, L.; Berkowitz, M. L.; Darden, T.; Lee, H.; Pedersen, L. G., A Smooth Particle Mesh Ewald Method. *J. Chem. Phys.* **1995,** *103* (19), 8577-8593.
77. Steinbrecher, T.; Mobley, D. L.; Case, D. A., Nonlinear scaling schemes for Lennard-Jones interactions in free energy calculations. *J. Chem. Phys.* **2007,** *127* (21), 214108.
78. Steinbrecher, T.; Joung, I.; Case, D. A., Soft-core potentials in thermodynamic integration: comparing one- and two-step transformations. *J. Comput. Chem.* **2011,** *32* (15), 3253-63.
79. Shirts, M. R.; Chodera, J. D., Statistically optimal analysis of samples from multiple equilibrium states. *J. Chem. Phys.* **2008,** *129* (12).
80. Bennett, C. H., Efficient Estimation of Free-Energy Differences from Monte-Carlo Data. *J. Comput. Phys.* **1976,** *22* (2), 245-268.
81. Roe, D. R.; Cheatham, T. E., PTRAJ and CPPTRAJ: Software for Processing and Analysis of Molecular Dynamics Trajectory Data. *J. Chem. Theory Comput.* **2013,** *9* (7), 3084-3095.
82. van der Walt, S.; Colbert, S. C.; Varoquaux, G., The NumPy Array: A Structure for Efficient Numerical Computation. *Comput. Sci. Eng.* **2011,** *13* (2), 22-30.
83. Roux, B.; Nina, M.; Pomes, R.; Smith, J. C., Thermodynamic stability of water molecules in the bacteriorhodopsin proton channel: A molecular dynamics free energy perturbation study. *Biophys. J.* **1996,** *71* (2), 670-681.
84. Gilson, M. K.; Given, J. A.; Bush, B. L.; McCammon, J. A., The statistical-thermodynamic basis for computation of binding affinities: A critical review. *Biophys. J.* **1997,** *72* (3), 1047-1069.
85. Mann, G.; Hermans, J., Modeling protein-small molecule interactions: Structure and thermodynamics of noble gases binding in a cavity in mutant phage T4 lysozyme L99A. *J. Mol. Biol.* **2000,** *302* (4), 979-989.
86. Boresch, S.; Tettinger, F.; Leitgeb, M.; Karplus, M., Absolute binding free energies: A quantitative approach for their calculation. *J. Phys. Chem. B* **2003,** *107* (35), 9535-9551.
87. Mobley, D. L.; Chodera, J. D.; Dill, K. A., On the use of orientational restraints and symmetry corrections in alchemical free energy calculations. *J. Chem. Phys.* **2006,** *125* (8).
88. Davis, M. E.; Mccammon, J. A., Electrostatics in Biomolecular Structure and Dynamics. *Chem. Rev.* **1990,** *90* (3), 509-521.

89. Sharp, K. A.; Honig, B., Electrostatic Interactions In Macromolecules - Theory And Applications. *Annu. Rev. Biophys. Biophys. Chem.* **1990,** *19*, 301-332.
90. Jeancharles, A.; Nicholls, A.; Sharp, K.; Honig, B.; Tempczyk, A.; Hendrickson, T. F.; Still, W. C., Electrostatic Contributions To Solvation Energies - Comparison Of Free-Energy Perturbation And Continuum Calculations. *J. Am. Chem. Soc.* **1991,** *113* (4), 1454-1455.
91. Honig, B.; Nicholls, A., Classical Electrostatics in Biology and Chemistry. *Science* **1995,** *268* (5214), 1144-1149.
92. Gilson, M. K., Theory Of Electrostatic Interactions In Macromolecules. *Curr. Opin. Struct. Biol.* **1995,** *5* (2), 216-223.
93. Beglov, D.; Roux, B., Solvation of complex molecules in a polar liquid: An integral equation theory. *J. Chem. Phys.* **1996,** *104* (21), 8678-8689.
94. Edinger, S. R.; Cortis, C.; Shenkin, P. S.; Friesner, R. A., Solvation free energies of peptides: Comparison of approximate continuum solvation models with accurate solution of the Poisson-Boltzmann equation. *J. Phys. Chem. B* **1997,** *101* (7), 1190-1197.
95. Cramer, C. J.; Truhlar, D. G., Implicit solvation models: Equilibria, structure, spectra, and dynamics. *Chem. Rev.* **1999,** *99* (8), 2161-2200.
96. Bashford, D.; Case, D. A., Generalized born models of macromolecular solvation effects. *Annu. Rev. Phys. Chem.* **2000,** *51*, 129-152.
97. Baker, N. A., Improving implicit solvent simulations: a Poisson-centric view. *Curr. Opin. Struct. Biol.* **2005,** *15* (2), 137-143.
98. Chen, J. H.; Im, W. P.; Brooks, C. L., Balancing solvation and intramolecular interactions: Toward a consistent generalized born force field. *J. Am. Chem. Soc.* **2006,** *128* (11), 3728-3736.
99. Feig, M.; Chocholousova, J.; Tanizaki, S., Extending the horizon: towards the efficient modeling of large biomolecular complexes in atomic detail. *Theor. Chem. Acc.* **2006,** *116* (1-3), 194-205.
100. Im, W.; Chen, J. H.; Brooks, C. L., Peptide and protein folding and conformational equilibria: Theoretical treatment of electrostatics and hydrogen bonding with implicit solvent models. *Peptide Solvation and H-Bonds* **2006,** *72*, 173-+.
101. Lu, B. Z.; Zhou, Y. C.; Holst, M. J.; McCammon, J. A., Recent progress in numerical methods for the Poisson-Boltzmann equation in biophysical applications. *Commun. Comput. Phys.* **2008,** *3* (5), 973-1009.
102. Wang, J.; Tan, C. H.; Tan, Y. H.; Lu, Q.; Luo, R., Poisson-Boltzmann solvents in molecular dynamics Simulations. *Commun. Comput. Phys.* **2008,** *3* (5), 1010-1031.
103. Altman, M. D.; Bardhan, J. P.; White, J. K.; Tidor, B., Accurate Solution of Multi-Region Continuum Biomolecule Electrostatic Problems Using the Linearized Poisson-Boltzmann Equation with Curved Boundary Elements. *J. Comput. Chem.* **2009,** *30* (1), 132-153.
104. Cai, Q.; Wang, J.; Hsieh, M.-J.; Ye, X.; Luo, R., Chapter Six - Poisson–Boltzmann Implicit Solvation Models. In *Annu. Rep. Comput. Chem.*, Ralph, A. W., Ed. Elsevier: 2012; Vol. Volume 8, pp 149-162.
105. Botello-Smith, W. M.; Cai, Q.; Luo, R., Biological applications of classical electrostatics methods. *J. Theor. Comput. Chem.* **2014,** *13* (03), 1440008.
106. Xiao, L.; Wang, C.; Luo, R., Recent progress in adapting Poisson–Boltzmann methods to molecular simulations. *J. Theor. Comput. Chem.* **2014,** *13* (03), 1430001.
107. Wen, E. Z.; Hsieh, M. J.; Kollman, P. A.; Luo, R., Enhanced ab initio protein folding simulations in Poisson-Boltzmann molecular dynamics with self-guiding forces. *J. Mol. Graphics Modell.* **2004,** *22* (5), 415-424.
108. Lwin, T. Z.; Zhou, R. H.; Luo, R., Is Poisson-Boltzmann theory insufficient for protein folding simulations? *J. Chem. Phys.* **2006,** *124* (3).

109. Wang, J.; Tan, C.; Chanco, E.; Luo, R., Quantitative analysis of Poisson-Boltzmann implicit solvent in molecular dynamics. *Phys. Chem. Chem. Phys.* **2010,** *12* (5), 1194-1202.
110. Wang, C.; Wang, J.; Cai, Q.; Li, Z.; Zhao, H.-K.; Luo, R., Exploring accurate Poisson–Boltzmann methods for biomolecular simulations. *Comput. Theor. Chem.* **2013,** *1024*, 34-44.
111. Wang, C. H.; Ren, P. Y.; Luo, R., Ionic Solution: What Goes Right and Wrong with Continuum Solvation Modeling. *J. Phys. Chem. B* **2017,** *121* (49), 11169-11179.
112. Case, D. A.; Cheatham, T. E.; Darden, T.; Gohlke, H.; Luo, R.; Merz, K. M.; Onufriev, A.; Simmerling, C.; Wang, B.; Woods, R. J., The Amber biomolecular simulation programs. *J. Comput. Chem.* **2005,** *26* (16), 1668-1688.
113. Tan, C.; Tan, Y. H.; Luo, R., Implicit nonpolar solvent models. *J. Phys. Chem. B* **2007,** *111* (42), 12263-74.
114. Botello-Smith, W. M.; Luo, R., Applications of MMPBSA to Membrane Proteins I: Efficient Numerical Solutions of Periodic Poisson-Boltzmann Equation. *J. Chem. Inf. Model.* **2015,** *55* (10), 2187-99.
115. Cai, Q.; Hsieh, M. J.; Wang, J.; Luo, R., Performance of Nonlinear Finite-Difference Poisson-Boltzmann Solvers. *J. Chem. Theory Comput.* **2010,** *6* (1), 203-211.
116. Wang, J.; Luo, R., Assessment of linear finite-difference Poisson-Boltzmann solvers. *J. Comput. Chem.* **2010,** *31* (8), 1689-98.
117. Lu, Q.; Luo, R., A Poisson–Boltzmann dynamics method with nonperiodic boundary condition. *J. Chem. Phys.* **2003,** *119* (21), 11035-11047.
118. Liptak, M. D.; Gross, K. C.; Seybold, P. G.; Feldgus, S.; Shields, G. C., Absolute pK(a) determinations for substituted phenols. *J. Am. Chem. Soc.* **2002,** *124* (22), 6421-6427.
119. Warshel, A.; Papazyan, A., Electrostatic effects in macromolecules: fundamental concepts and practical modeling. *Curr. Op. Struct. Biol.* **1998,** *8* (2), 211-217.
120. Schutz, C. N.; Warshel, A., What are the dielectric "constants" of proteins and how to validate electrostatic models? *Proteins: Struct., Func., and Bioinf.* **2001,** *44* (4), 400-417.
121. Gouda, H.; Kuntz, I. D.; Case, D. A.; Kollman, P. A., Free energy calculations for theophylline binding to an RNA aptamer: comparison of MM‐PBSA and thermodynamic integration methods. *Biopolymers* **2003,** *68* (1), 16-34.
122. Kollman, P. A.; Massova, I.; Reyes, C.; Kuhn, B.; Huo, S.; Chong, L.; Lee, M.; Lee, T.; Duan, Y.; Wang, W., Calculating structures and free energies of complex molecules: combining molecular mechanics and continuum models. *Acc. Chem. Res.* **2000,** *33* (12), 889-897.
123. Bezencon, J.; Wittwer, M. B.; Cutting, B.; Smiesko, M.; Wagner, B.; Kansy, M.; Ernst, B., pK(a) determination by H-1 NMR spectroscopy - An old methodology revisited. *J. Pharm. Biomed. Anal.* **2014,** *93*, 147-155.
124. Khare, D.; Alexander, P.; Antosiewicz, J.; Bryan, P.; Gilson, M.; Orban, J., pKa Measurements from Nuclear Magnetic Resonance for the B1 and B2 Immunoglobulin G-Binding Domains of Protein G: Comparison with Calculated Values for Nuclear Magnetic Resonance and X-ray Structures. *Biochemistry* **1997,** *36* (12), 3580-3589.
125. Papaneophytou, C. P.; Grigoroudis, A. I.; McInnes, C.; Kontopidis, G., Quantification of the Effects of Ionic Strength, Viscosity, and Hydrophobicity on Protein-Ligand Binding Affinity. *ACS Med. Chem. Lett.* **2014,** *5* (8), 931-936.
126. Beierlein, F. R.; Michel, J.; Essex, J. W., A simple QM/MM approach for capturing polarization effects in protein-ligand binding free energy calculations. *J. Phys. Chem. B* **2011,** *115* (17), 4911-26.
127. Olsson, M. A.; Ryde, U., Comparison of QM/MM Methods To Obtain Ligand-Binding Free Energies. *J. Chem. Theory Comput.* **2017,** *13* (5), 2245-2253.
128. Lamoureux, G.; Roux, B., Modeling induced polarization with classical Drude oscillators: Theory and molecular dynamics simulation algorithm. *J. Chem. Phys.* **2003,** *119* (6), 3025-3039.

129. Aguilar, B.;  Anandakrishnan, R.;  Ruscio, J. Z.; Onufriev, A. V., Statistics and Physical Origins of pK and Ionization State Changes upon Protein-Ligand Binding. *Biophys. J.* **2010,** *98* (5), 872-880.
130. Bas, D. C.;  Rogers, D. M.; Jensen, J. H., Very fast prediction and rationalization of pK(a) values for protein-ligand complexes. *Proteins: Struct., Func., and Bioinf.* **2008,** *73* (3), 765-783.
131. Mackerell, A. D.;  Sommer, M. S.; Karplus, M., Ph-Dependence of Binding Reactions from Free-Energy Simulations and Macroscopic Continuum Electrostatic Calculations - Application to 2'gmp/3'gmp Binding to Ribonuclease T-1 and Implications for Catalysis. *J. Mol. Bio.* **1995,** *247* (4), 774-807.
132. Onufriev, A. V.; Alexov, E., Protonation and pK changes in protein-ligand binding. *Q. Rev. Biophys.* **2013,** *46* (2), 181-209.
133. Kim, M. O.; McCammon, J. A., Computation of pH-dependent binding free energies. *Biopolymers* **2016,** *105* (1), 43-49.
134. Chen, W.;  Morrow, B. H.;  Shi, C. Y.; Shen, J. K., Recent development and application of constant pH molecular dynamics. *Mol. Simul.* **2014,** *40* (10-11), 830-838.

2727

# Supplementary Information

# Estimating the Roles of Protonation and Electronic Polarization in Absolute Binding Affinity Simulations

*Edward King*[1], *Ruxi Qi*[4], *Han Li*[2], *Ray Luo*[1,2,3*], *and Erick Aitchison*[1*]

[1]Departments of Molecular Biology and Biochemistry, [2]Chemical and Biomolecular Engineering, [3]Materials Science and Engineering and Biomedical Engineering, University of California, Irvine, California 92697 (USA)
[4]Cryo-EM Center, Southern University of Science and Technology, Shenzhen, Guangdong 518055 (China)

## SI Discussion

Apparent Binding Free Energies with One Titratable Group in the Active Site

The presence of titratable groups in the active site poses an additional challenge for the calculation of binding free energies. The experimentally resolved apparent binding free energies encapsulate the whole physical process, which may not distinguish the contributions of coupled processes to the protein-ligand binding event. Titratable groups in the active site are susceptible to the system pH, and this susceptibility is observed in the pH dependence of receptor-ligand binding[1,2] and enzymatic catalysis[3]. This can be further complicated when the interaction of the binding ligand shifts the pKa of those titratable groups which can alter protonation states. These interactions and coupled processes need to be considered for the binding free energy calculations. Using a single titratable group as a model coupled process, one can derive the separable contributions from the coupled processes. The coupled binding process can be separated into four separate processes: binding in the protonated form, binding in the deprotonated form, and protonation/deprotonation processes in the free and complex states defined in the illustrated thermodynamic cycle where one of the binding processes is directly calculated.

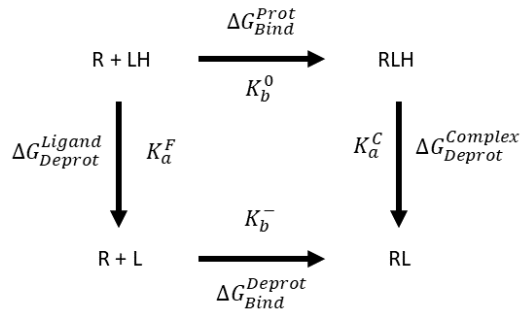

The apparent binding constant can be expressed with a proton dissociation process for both the complex and free states,

$$K_{app} = \frac{[RLH] + [RL^-]}{[R][LH] + [R][L^-]}$$

This can be further simplified by substitution to give,

$$K_{app} = K_b^0 \frac{(1 + (K_a^C)^{-1}[H])}{(1 + (K_a^F)^{-1}[H])}$$

where $K_b^0 = \frac{[RLH]}{[R][LH]}$ is the protonated binding equilibrium constant, $K_a^C = \frac{[RL^-][H]}{[RLH]}$ and $K_a^F = \frac{[L^-][H]}{[LH]}$ are the proton dissociation constants in the complex and free states, respectively. The proton dissociation constants can be expressed in pKa and pH units, and the change in free energy can be calculated using

$$\Delta G^o(pH) = -k_b T \left[\ln K_b^0 + \ln \frac{(1+10^{pH-pKa^C})}{(1+10^{pH-pKa^F})}\right]^{4,\,5}$$

The charged binding equilibrium constant can be converted to the binding free energy which can be explicitly calculated, where $\Delta G_{bind}^{Prot} = -k_b T \ln K_b^0$, resulting in

$$\Delta G^o(pH) = \Delta G_{bind}^{Prot} - k_b T \ln \frac{(1 + 10^{pH-pKa^C})}{(1 + 10^{pH-pKa^F})}$$

Additionally, this equation then becomes process dependent, where the equation for binding coupled to a proton association process is,

$$\Delta G^o(pH) = \Delta G_{bind}^{Deprot} - k_b T \ln \frac{(1 + 10^{pKa^C - pH})}{(1 + 10^{pKa^F - pH})}$$

Application of the above equation shows that computation of the apparent binding free energy requires the pKa's of the ligand in both the free and bound states in addition to the simulated binding affinity of the ligand in either of the states. However, the application of this simplified single titratable group coupled binding process equation is apparently inadequate in describing the complete binding process for most protein systems where many residues in the active site are also titratable and require proper modeling.

**SI Tables**

| Condition | Epsin | RMSE (kcal/mol) | R |
|---|---|---|---|
| All HIP | 1 | 2.31 | 0.60 |
| All HID | 1 | 3.75 | 0.75 |
| Small HIP | 1 | 2.89 | 0.85 |
| Small HID | 1 | 3.32 | 0.17 |

**Table 1. MBAR/PBSA binding affinity accuracy with optimized Radiscale and Protscale parameters.** Radiscale and Protscale values were scaled to minimize mean absolute error between MBAR/PBSA free energies and explicit solvent free energies.

| Condition | Epsin | RMSE (kcal/mol) | R |
|---|---|---|---|
| All-HIP | 1.17 | 1.25 | 0.74 |
| All-HIP | 2 | 4.84 | 0.81 |
| All-HID | 1.43 | 0.89 | 0.88 |
| All-HID | 2 | 2.65 | 0.88 |
| Small-HIP | 1.27 | 1.00 | 0.88 |
| Small-HIP | 2 | 3.91 | 0.87 |
| Small-HID | 1.28 | 1.80 | 0.32 |
| Small-HID | 2 | 3.90 | 0.52 |

**Table 2. Binding affinity prediction accuracy versus solute interior dielectric (Epsin) with MBAR/PBSA.** Commonly used Epsin of 2.0 and Epsin resulting in the lowest RMSE to experiment are reported. All-HID condition shows the lowest RMSE and highest Pearson correlation at optimized Epsin. Epsin 2.0 results in improved Pearson correlations, but also higher RMSE's.

| Sample | Baseline | Baseline + 150mM salt | Baseline + deprotonated ligands | 1DOF All-Hip | 6DOF All-Hip | All HID | Small HIP | Small HID | PBSA All HIP | PBSA All HID | PBSA Small HIP | PBSA Small HID | Experiment |
|---|---|---|---|---|---|---|---|---|---|---|---|---|---|
| 1C5X | -13.32 | -11.15 | -13.32 | -11.15 | -14.29 | -12.68 | -11.15 | -12.68 | -8.81 | -8.55 | -7.78 | -9.68 | -9.01 |
| 1C5Y | -10.83 | -10.37 | -10.83 | -10.37 | -13.91 | -11.66 | -10.37 | -11.66 | -7.47 | -6.93 | -6.35 | -8.09 | -5.67 |
| 1C5Z | -9.89 | -6.87 | -9.89 | -6.87 | -8.95 | -10.74 | -6.87 | -10.74 | -4.60 | -6.60 | -3.57 | -7.70 | -5.42 |
| 1GI7 | -8.88 | -9.68 | -8.88 | -9.68 | -11.39 | -10.26 | -9.68 | -10.26 | -7.93 | -6.81 | -7.21 | -7.73 | -6.09 |
| 1GJ7 | -6.87 | -7.35 | -13.10 | -12.08 | -17.74 | -13.41 | -13.41 | -12.08 | -10.12 | -10.49 | -11.23 | -9.19 | -10.86 |
| 1GJ8 | -8.35 | -6.89 | -7.92 | -7.99 | -13.66 | -11.51 | -11.51 | -7.99 | -7.13 | -7.86 | -8.77 | -6.71 | -9.39 |
| 1GJA | -8.49 | -8.15 | -6.78 | -7.82 | -9.15 | -10.87 | -10.87 | -7.82 | -6.32 | -6.38 | -7.60 | -5.54 | -7.32 |
| 1GJB | -8.64 | -8.44 | -9.89 | -11.23 | -15.75 | -12.50 | -12.50 | -11.23 | -8.56 | -8.58 | -9.59 | -7.47 | -8.57 |
| 1GJD | -4.05 | -5.09 | -5.47 | -5.08 | -9.14 | -4.85 | -4.85 | -5.08 | -3.60 | -3.03 | -3.48 | -2.73 | -7.05 |
| 1O3P | -7.66 | -10.36 | -12.63 | -11.41 | -16.18 | -12.73 | -12.73 | -11.41 | -9.19 | -8.81 | -9.86 | -8.24 | -8.99 |

**Table 3. Full binding predictions at all conditions tested compared to experimental values.** Absolute binding free energy calculations aggregated from 5 independent replicates with randomized starting velocities. All units reported in kcal/mol.

# SI Figures

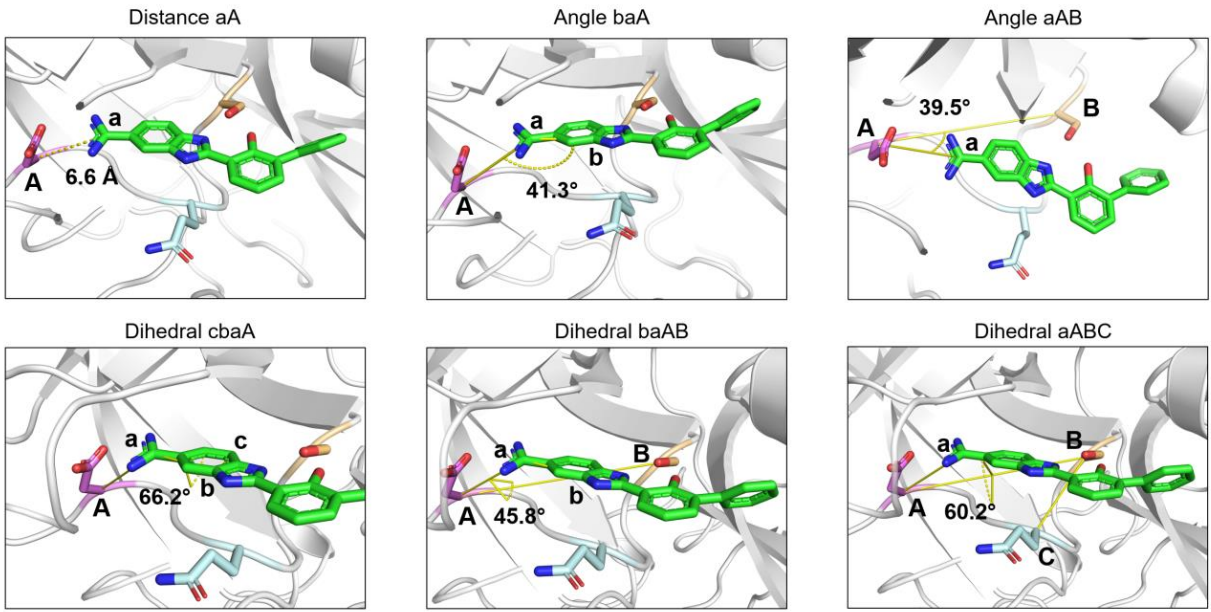

**Figure 1. Illustration of Boresch 6DOF orientational restraints.** The ligand is constrained by a single distance, two angles, and three dihedrals selected from the end of the equilibration phase to lock the ligand into a target conformation. 1DOF condition involves only the distance restraint, which allows greater exploration of conformational states at the cost of slower convergence.

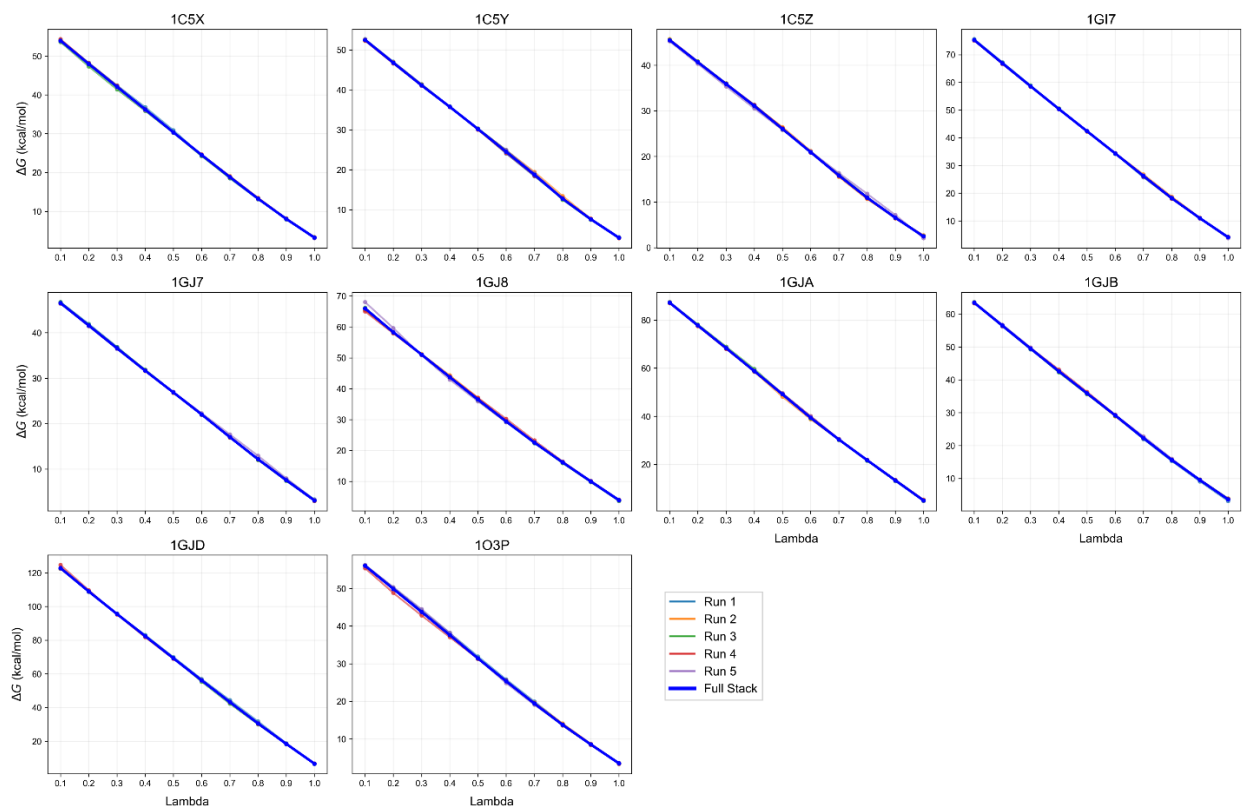

**Figure 2. Free energy transitions during the decharging phase for the complex trajectories in the baseline simulation.** Individual replicates show only small variation, the aggregated energies show almost complete overlap and smooth, nearly linear transition from full ligand partial charges to zero.

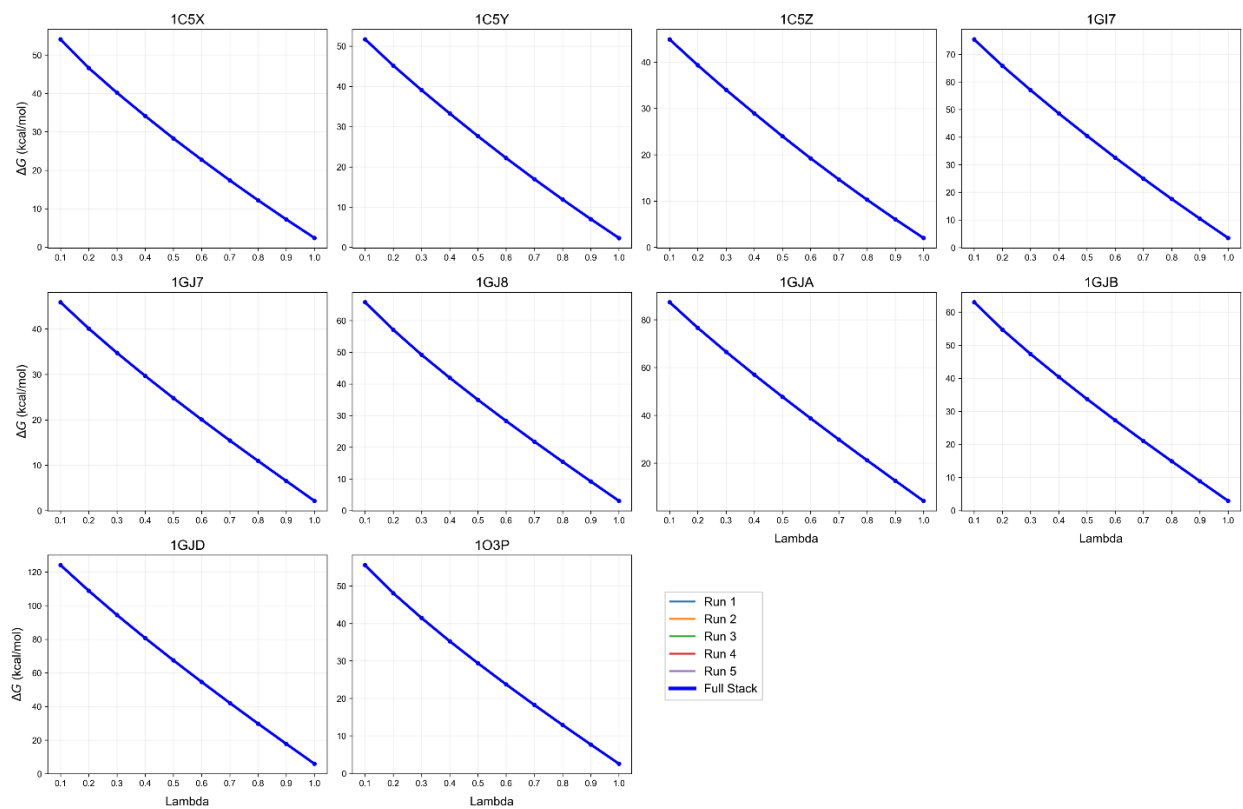

**Figure 3. Free energy transition during the decharging phase for the ligand trajectories in the baseline alchemical simulation.** The same pattern of small variation and linear transition from full ligand partial charges to zero as the complex is observed.

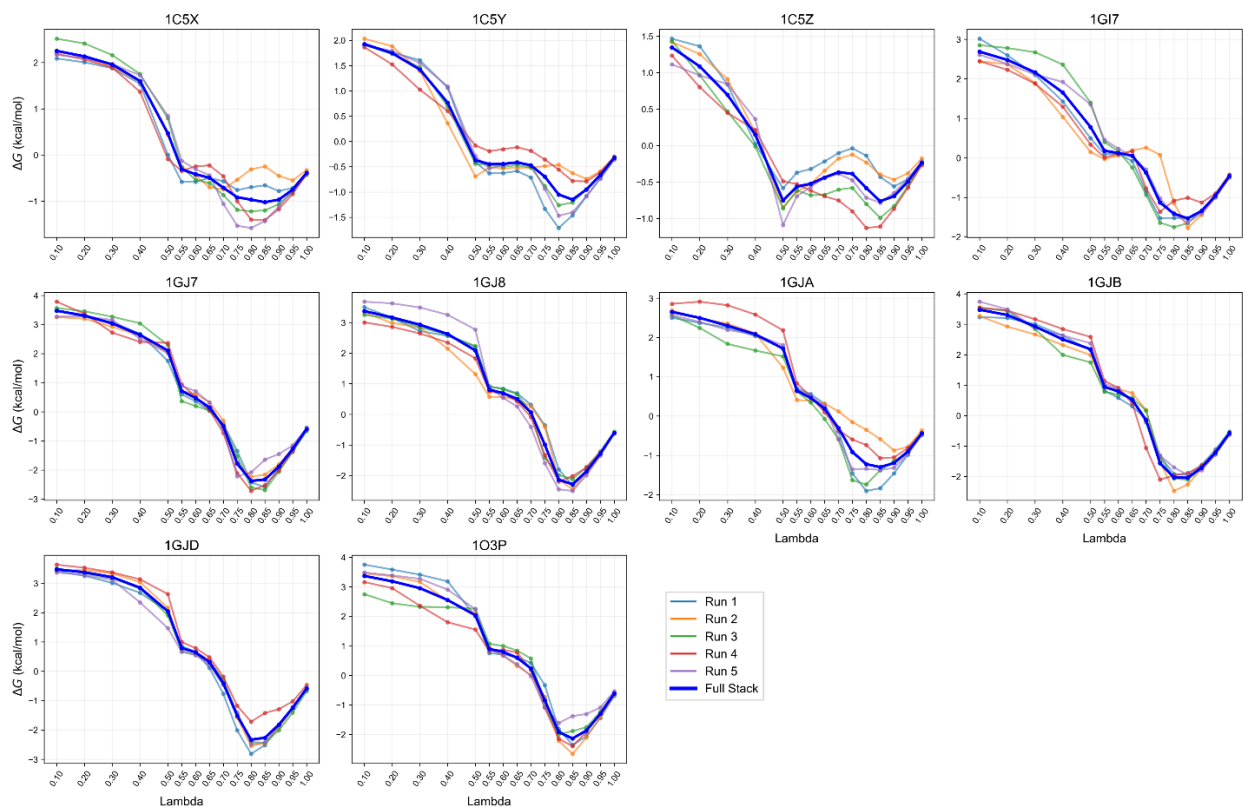

**Figure 4. Free energy transition during the VDW phase for the complex trajectories in the baseline simulation.** High variance is observed between replicates, highlighting the sampling difficulties associated with decoupling VDW interactions.

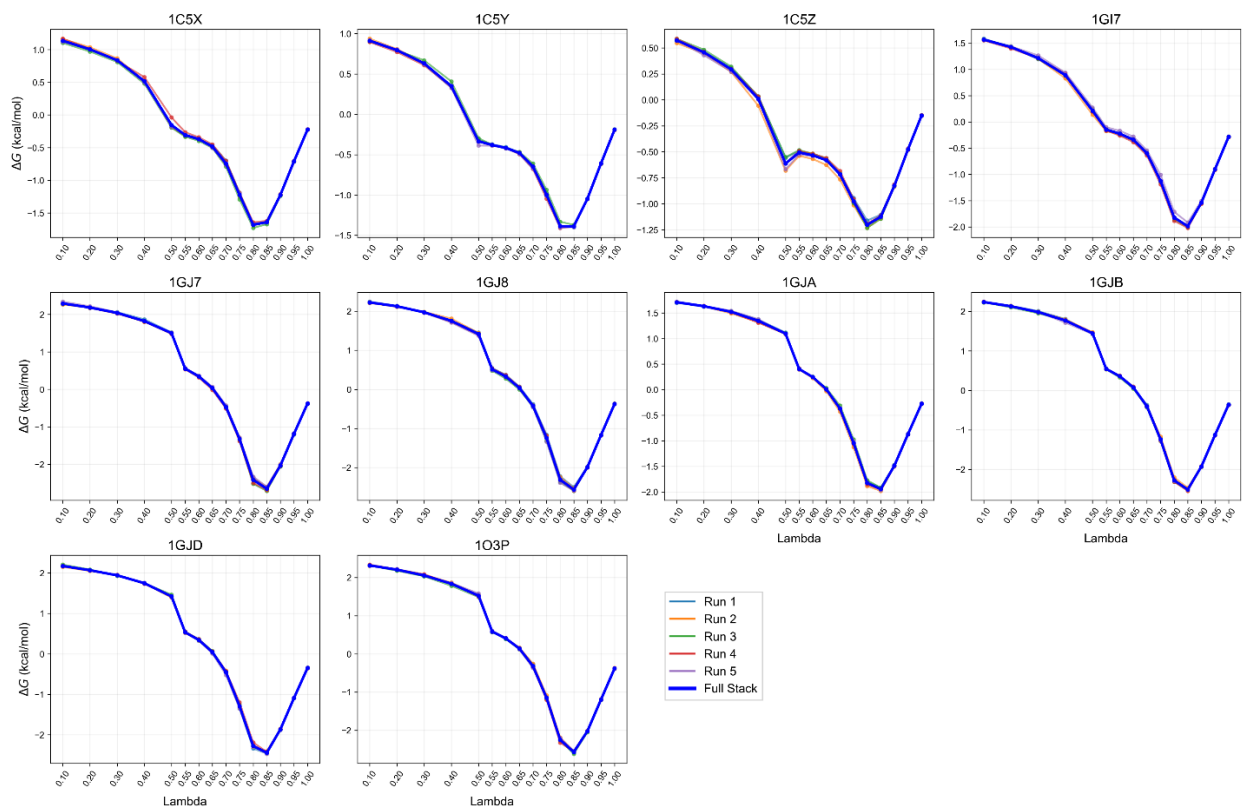

**Figure 5. Free energy transition during the VDW phase for the ligand trajectories in the baseline simulation.** Replicates show high agreement over the course of the highly non-linear transitions.

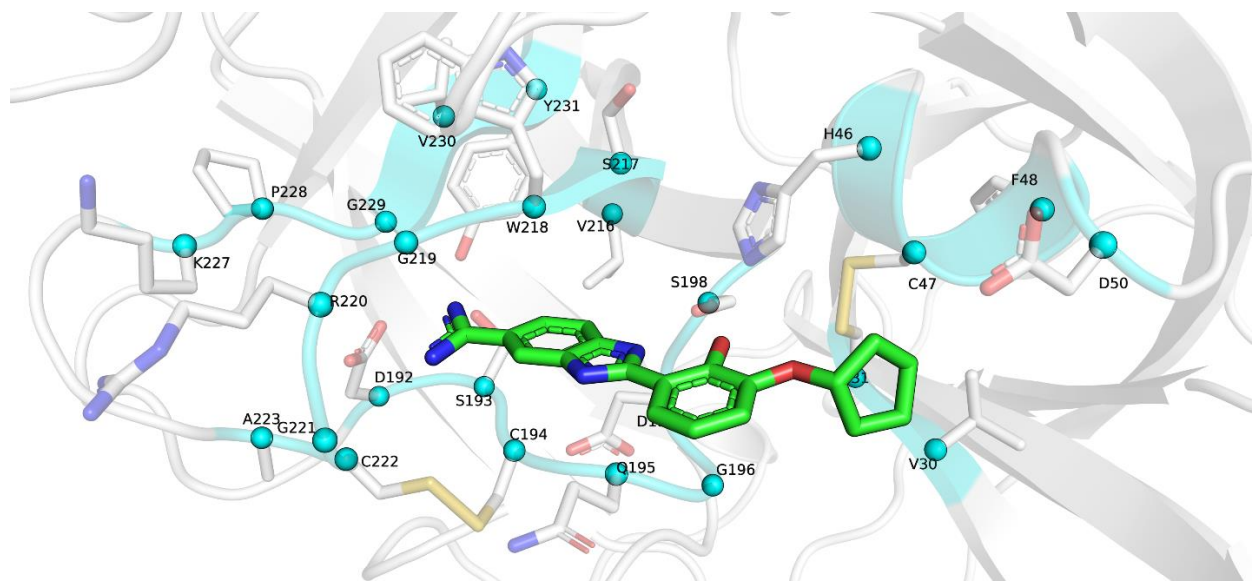

**Figure 6. Illustration of the UPA binding pocket with all residues within 6 Å of the ligand highlighted.** Notable residues include His-46 which is titratable and observed to form a hydrogen bond with the ligand phenol. Asp-192 is located at the base of the binding pocket and forms salt bridges with the positively charged amidine. Sample ligand 1O3P is highlighted in green.

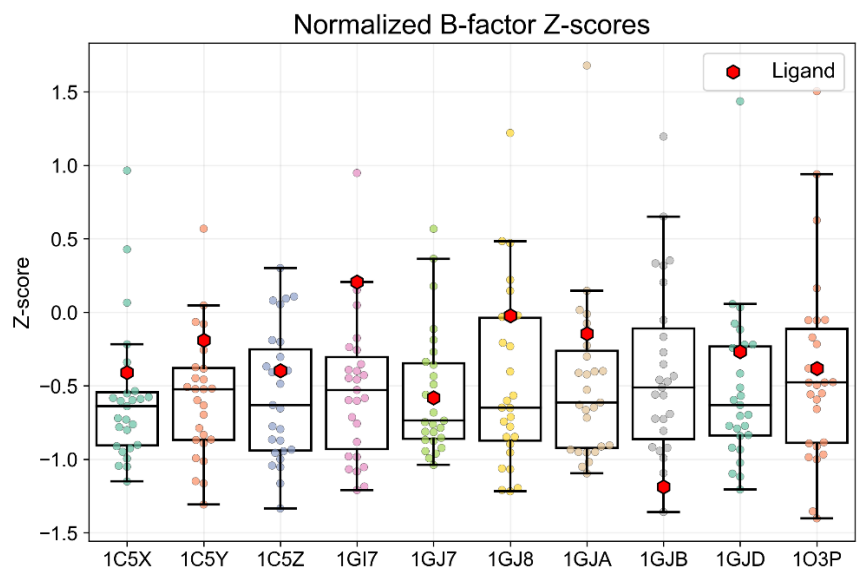

**Figure 7. Analysis of binding pocket flexibility through normalized B-factor Z-scores.** All structures show similar binding pocket flexibility, with higher than average rigidity relative to the rest of the protein. Ligands show varying levels of displacement, notably 1GI7 shows the highest flexibility, which is larger in size but unable to form a hydrogen bond to Ser-198. 1GJB shows the highest stability, potentially due to its hydrophobic benzene groups and internal hydrogen bond between the ligand phenol and nitrogen. Each marker represents the Z-score per residue with all atoms averaged.

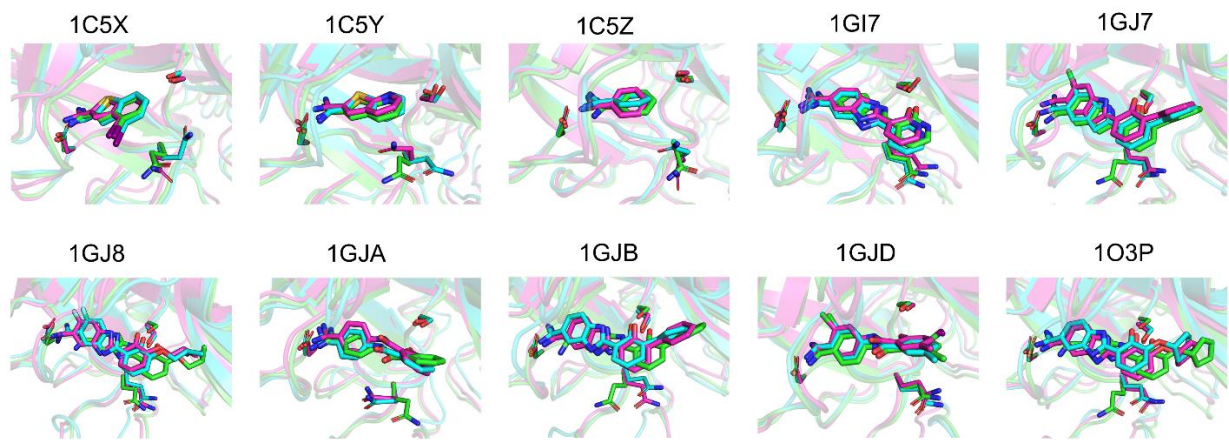

**Figure 8. Inhibitor equilibration poses from GAFF and GAFF2 compared to starting crystal poses.** GAFF and GAFF2 trajectories show similar trends, with the ligands moving further into the binding pocket to more tightly interact with Asp-192, and outward twisting of the phenol tail to relieve steric clash. Structures were generated from identifying the frame with the lowest RMSD to the average structure from the last 10 ns of equilibration. The starting crystal structure models are colored green, GAFF samples are colored cyan, and GAFF2 samples are colored purple.

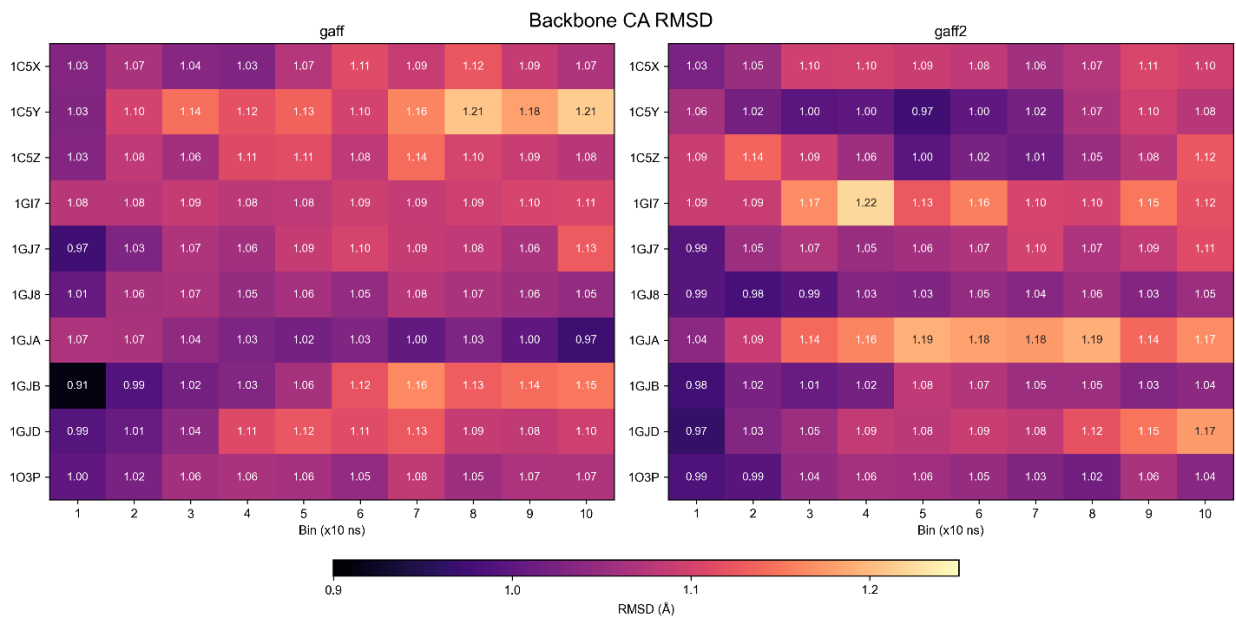

**Figure 9. Backbone CA RMSD development over equilibration with GAFF and GAFF2 force fields.** No clear pattern emerges, all proteins drift away from the starting ligand pose and show a maximum divergence of ~1.2 Å RMSD, indicating that minor conformational changes occur.

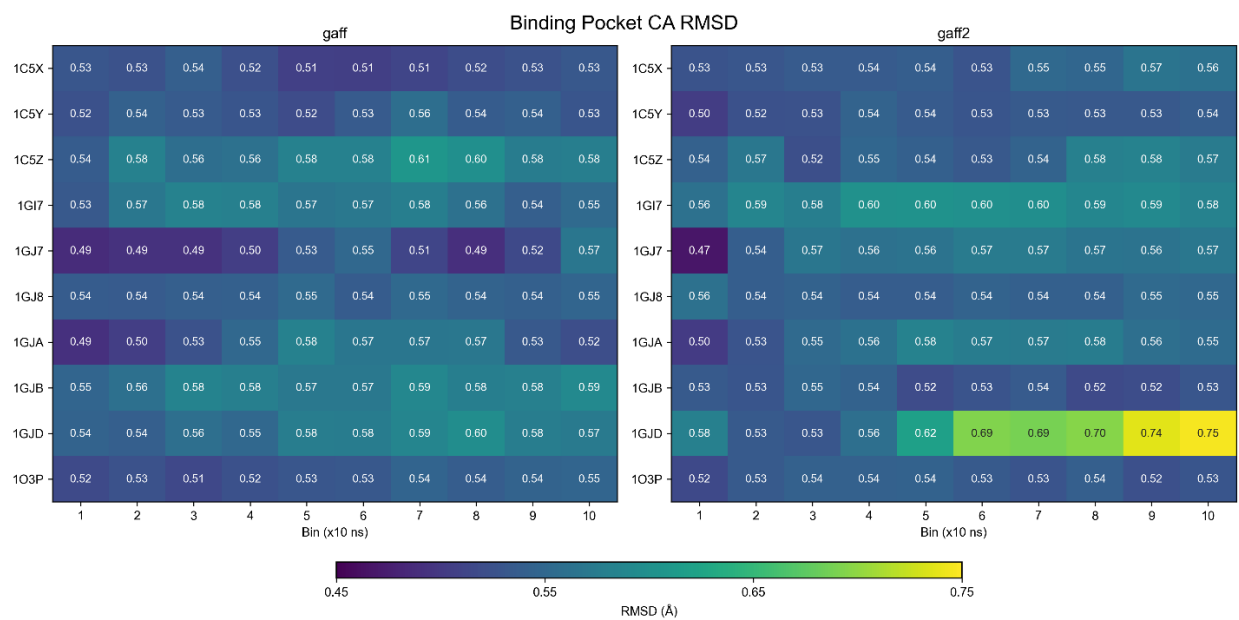

**Figure 10. Binding pocket CA RMSD development over equilibration with GAFF and GAFF2 force fields.** All GAFF samples show stability and do not change noticeably from the crystal over the course of equilibration. In GAFF2, 1GJD shows larger divergence from the crystal pose reaching 0.75 A RMSD.

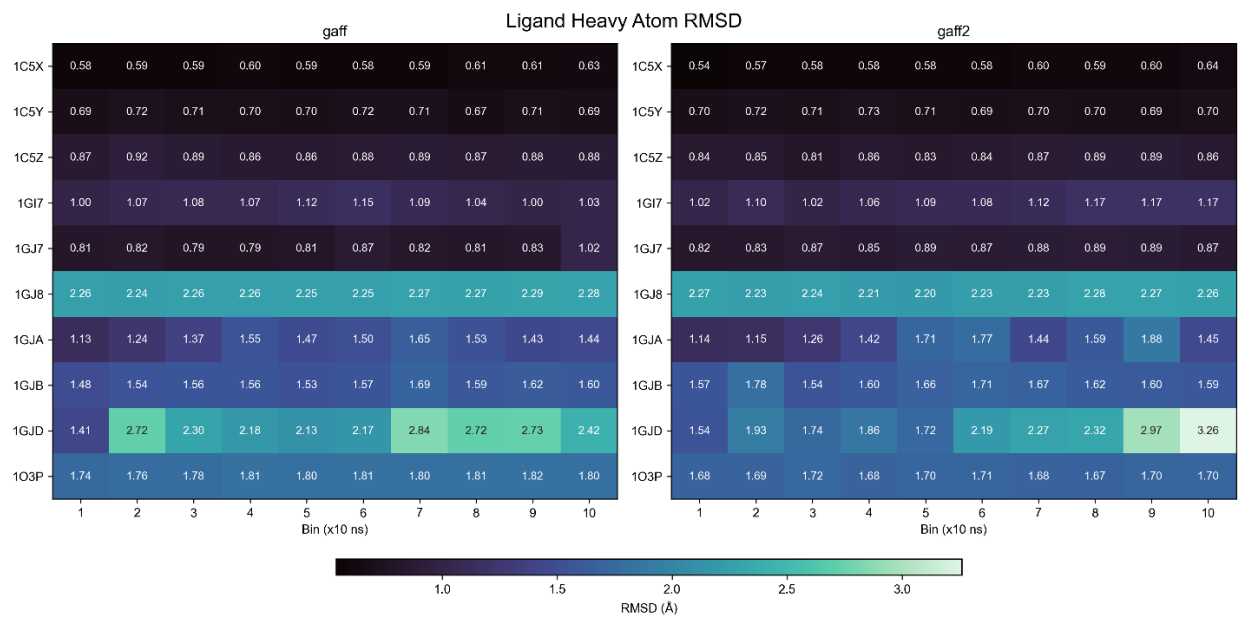

**Figure 11. Ligand heavy atom RMSD development over the equilibration with GAFF and GAFF2 force fields.** Small ligands (1C5X, 1C5Y, 1C5Z, and 1GI7) show minimal changes in positioning. 1GJ8 shows consistent departure from the crystal pose, the ligand moves further into the binding pocket to maximize hydrophobic interactions and polar interactions with Asp-192. 1GJD shows dissimilarity with crystal as well, from the rotation of the phenol group outward leading to the loss of the hydrogen bond. The aberration is more substantial with GAFF2.

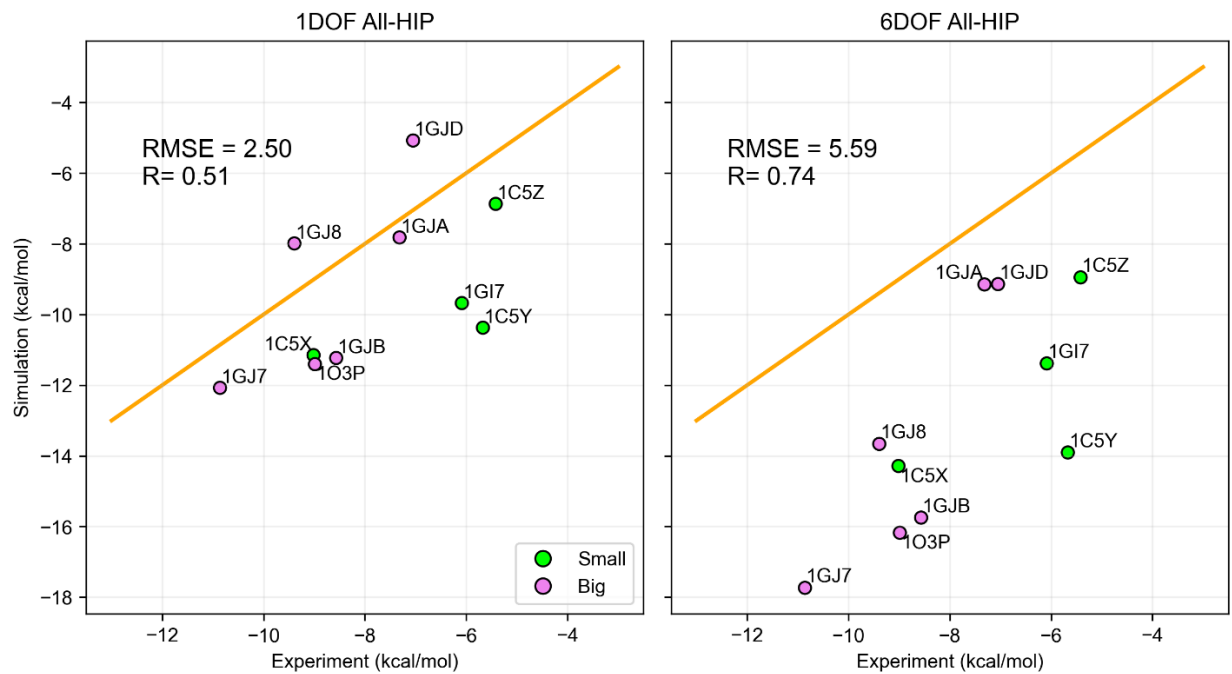

**Figure 12. Comparison of 1DOF and 6DOF restraint schemes.** The 1DOF single distance restraint showed lower error, but worse Pearson correlation than the 6DOF (Boresch) method. Samples with the 6DOF restraint showed excessively negative free energy predictions, indicating potential over-stabilization in a favorable pose.

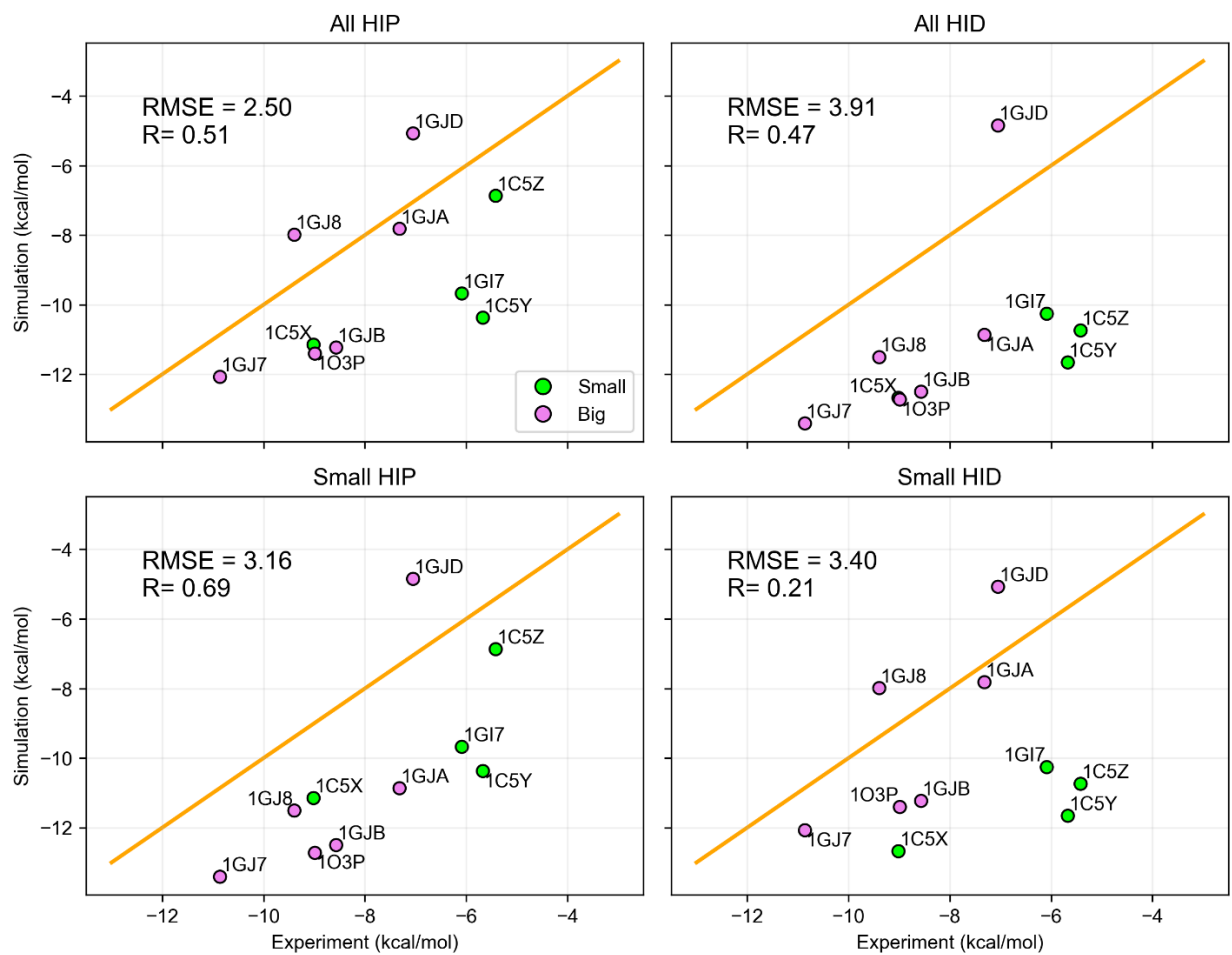

**Figure 13. Binding affinity predictions with standard alchemical simulation with different protonation states.** In general, binding affinities are predicted to be more negative than expected, possibly due to exaggeration of favorable charge-charge interactions typical of the point-charge models used. 1GJD is shown to be an outlier, with free energies far more positive than the cluster of other tested ligands, this is likely related to the issues in sampling incorrect binding poses recognized during equilibration where the phenol swings outward such that the native hydrogen bond to Ser-198 is not maintained.

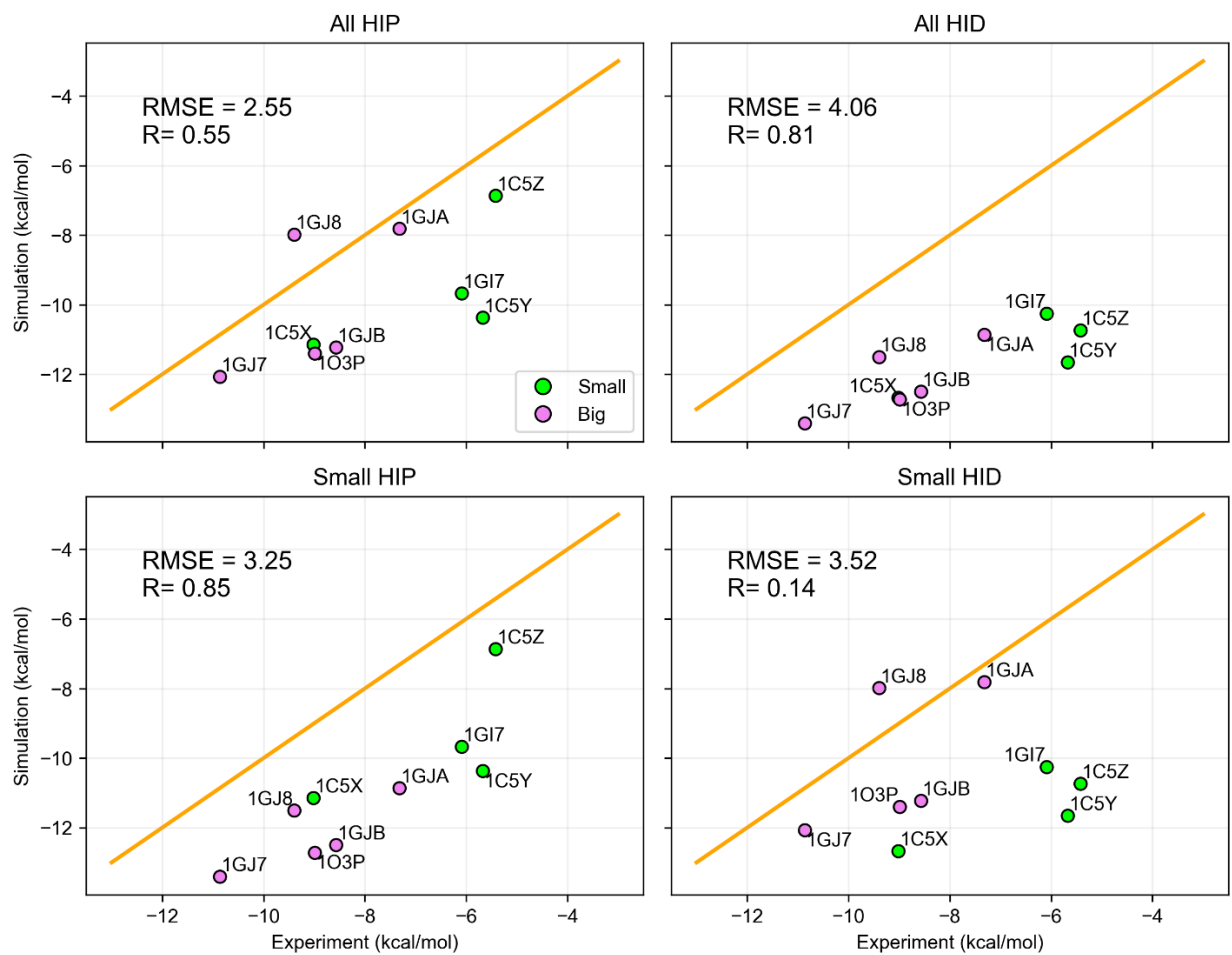

**Figure 14. Binding affinity predictions with outlier 1GJD removed for standard alchemical simulation with different protonation states.** In the standard alchemical simulation, minimal change is seen in RMSE for all conditions. However, Pearson correlation is found to improve dramatically for both All-HID and Small-HIP conditions.

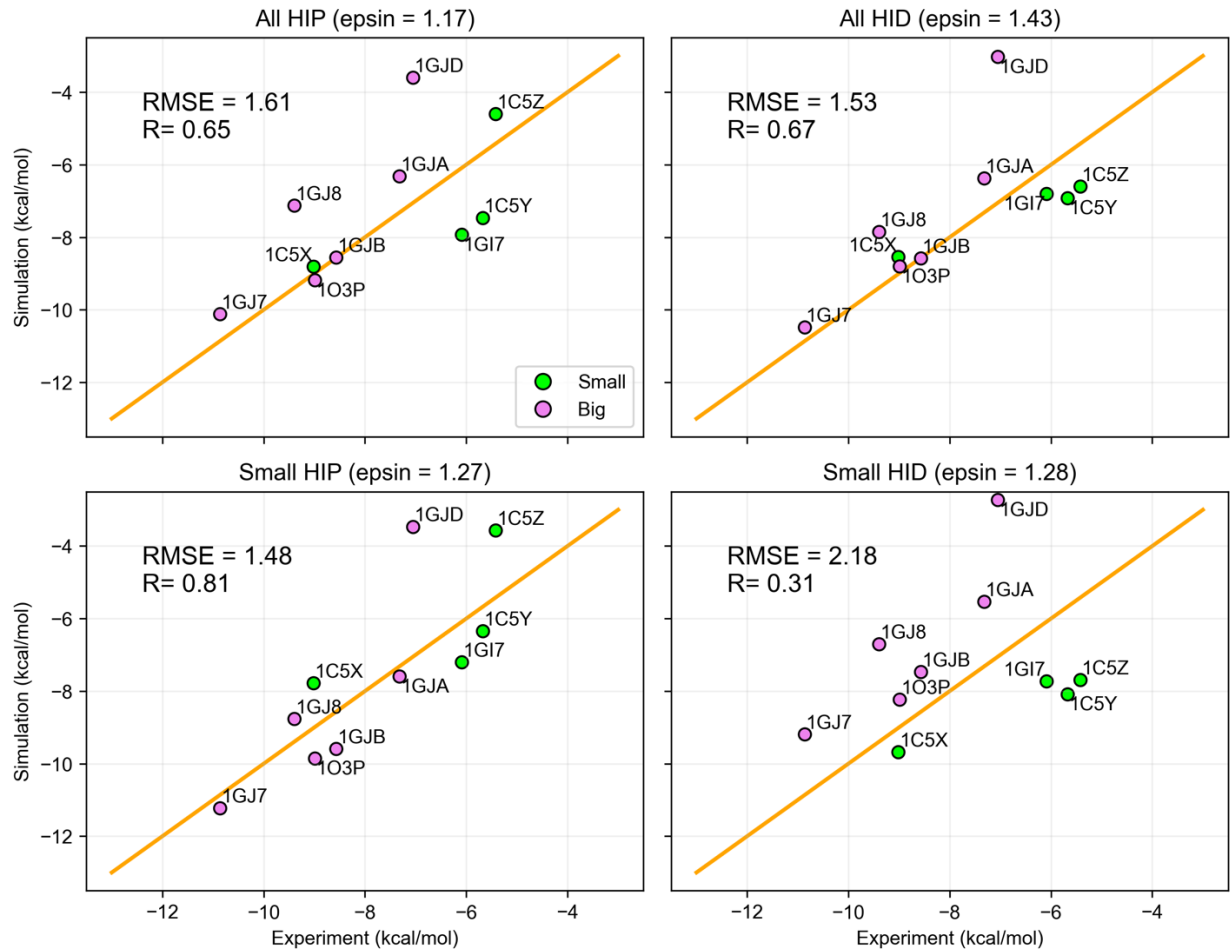

**Figure 15. MBAR/PBSA binding affinity calculations including the outlier 1GJD.** All metrics are found to worsen with the outlier pushing the trend toward overly positive values.

## SI References

1. Maiti, A.; Drohat, A. C., Dependence of substrate binding and catalysis on pH, ionic strength, and temperature for thymine DNA glycosylase: Insights into recognition and processing of G.T mispairs. *DNA Repair (Amst)* **2011,** *10* (5), 545-53.
2. Hansen, M. J.; Olsen, J. G.; Bernichtein, S.; O'Shea, C.; Sigurskjold, B. W.; Goffin, V.; Kragelund, B. B., Development of prolactin receptor antagonists with reduced pH-dependence of receptor binding. *J Mol Recognit* **2011,** *24* (4), 533-47.
3. Talley, K.; Alexov, E., On the pH-optimum of activity and stability of proteins. *Proteins* **2010,** *78* (12), 2699-706.
4. Jensen, J. H., Calculating pH and salt dependence of protein-protein binding. *Curr Pharm Biotechnol* **2008,** *9* (2), 96-102.
5. Mason, A. C.; Jensen, J. H., Protein-protein binding is often associated with changes in protonation state. *Proteins* **2008,** *71* (1), 81-91.